\documentclass{article}

\usepackage[utf8]{inputenc}   
\usepackage[T1]{fontenc}
\usepackage{lmodern}          
\usepackage{textcomp}         
\usepackage[hidelinks,pdfencoding=auto,psdextra]{hyperref}

\usepackage{amsmath,amssymb,amsthm,mathtools}

\usepackage{graphicx}
\graphicspath{{figures/}}
\usepackage{booktabs}         
\usepackage{multirow}
\usepackage{array}
\usepackage{tabularx}

\usepackage{xcolor}

\hypersetup{
  pdftitle={Generative Logic: A Deterministic Reasoning Engine},
  pdfauthor={Nikolai Sergeev},
  pdfkeywords={automated reasoning, formal methods, theorem discovery, knowledge generation, symbolic AI, hardware-software codesign}
}

\usepackage[nameinlink,capitalise,noabbrev]{cleveref}

\usepackage[a4paper,margin=1in]{geometry}

\usepackage{listings}
\lstdefinelanguage{MPL}{
  sensitive=true,
  morecomment=[l]{//},
  morecomment=[s]{/*}{*/},
  morestring=[b]"
}
\definecolor{mpl-bg}{rgb}{0.95,0.95,0.92}
\definecolor{mpl-kw}{rgb}{0.00,0.00,0.80}
\definecolor{mpl-cm}{rgb}{0.00,0.60,0.00}
\definecolor{mpl-str}{rgb}{0.00,0.00,0.60}

\lstdefinestyle{mplstyle}{
  language=MPL,
  backgroundcolor=\color{mpl-bg},
  basicstyle=\ttfamily\footnotesize,
  keywordstyle=\color{mpl-kw},
  commentstyle=\color{mpl-cm},
  stringstyle=\color{mpl-str},
  numbers=left,
  numbersep=5pt,
  breaklines=true,
  showstringspaces=false,
  tabsize=2,
  captionpos=b
}
\crefname{lstlisting}{Listing}{Listings}
\Crefname{lstlisting}{Listing}{Listings}

\usepackage[numbers,sort&compress]{natbib}

\title{Generative Logic: A Deterministic Reasoning Engine}

\author{Nikolai Sergeev\\
\small Generative Logic UG, Germany\\
\small\href{mailto:nikolai.sergeev@generative-logic.com}{\texttt{nikolai.sergeev@generative-logic.com}}
}

\date{\textbf{October 30, 2026}}

\begin{document}
\maketitle

\begin{abstract}
We present Generative Logic (GL), a deterministic computer architecture for automated mathematical reasoning. GL starts from axiomatic definitions written in a minimalist Mathematical Programming Language (MPL) and explores a configurable region of their deductive neighborhood. The definitions are compiled into a grid of Logic Blocks (LBs). An LB applies rules by hash-table lookups, and every statement it derives keeps its complete provenance. A run generates conjectures, removes those that counterexamples refute, tries to prove the rest, reduces the proved theorems to an essential core and writes every proof as navigable HTML. An independent verifier checks every proof step. From supplied definitions, GL derives the basic laws of Peano arithmetic and Gauss's summation formula. From previously proved lemmas, further definitions and a supplied list of conjectures, it builds a proof of the Fundamental Theorem of Arithmetic (FTA). These proofs are also exported to Lean~4 and checked by the Lean kernel. The next step is for GL to derive the FTA as well, without supplied conjectures. GL is designed together with a hardware architecture, the Mathematical Processing Unit (MPU).
\end{abstract}

\section{Introduction}
\label{sec:intro}

Automated mathematical reasoning follows two main approaches. Provers based on Large Language Models (LLMs) solve many formal competition problems; a human states each problem, and Lean checks the proof \cite{hubert2026alphaproof,ren2025deepseekproverv2}. Interactive proof assistants such as Lean \cite{demoura2021lean4} and Isabelle \cite{nipkow2002isabelle} check every proof mechanically, but writing the proofs needs considerable manual effort and expertise \cite{lu2024palm}. A third, smaller line of work generates conjectures from definitions and tries to prove them without human goal-setting (\cref{sec:related}).

Generative Logic (GL) belongs to this third line. GL is a computer architecture that takes axiomatic definitions written in a minimalist Mathematical Programming Language (MPL) and explores the deductive space they define: it generates conjectures, removes those that counterexamples refute, tries to prove all the others at once and records every proof step with its justification. The unit of work is therefore not a single theorem stated by a user but the family of theorems that follows from the definitions. GL is deterministic: a run repeated with the same input, configuration and processing route (processor or graphics processor, \cref{sec:mpu}) produces the same proofs. An independent verifier checks every proof step, and proofs can also be exported to Lean~4 and checked by the Lean kernel (\cref{sec:output}). A deterministic engine of this kind can also supply checked results to probabilistic AI systems, a combination that other work has proposed as well \cite{zhang2024fusion}.

GL can be understood through an analogy with biological development. A living organism does not store an explicit blueprint of every cell, tissue and organ it will produce. Instead, a compact set of genetic instructions, the DNA, encodes a small collection of foundational rules, and the machinery of the cell does the rest: molecules interact, differentiate and organize themselves into complex structures. This unfolding is not unbounded. Gene regulation, energy budgets and environmental signals determine which developmental paths are expressed in a given context; the genome encodes far more potential than any single organism realizes. GL applies the same principle to logic. A minimal set of axiomatic definitions, written in MPL, plays the role of the genome: a compact encoding of the foundations. The GL architecture acts as the developmental machinery, a deterministic engine that explores the deductive consequences of the definitions. The configuration files play a role similar to gene regulation: they do not change the logical content of the definitions, but they determine which region of the deductive space a run explores. Just as no biologist specifies every protein interaction for an organism to grow, no mathematician states the theorems of Peano arithmetic or Gauss's summation formula for GL to find them. The axioms are planted; the proof graph grows. With more computational resources, the constraints can be relaxed, which allows a broader exploration from the same definitions. This shift from explicit enumeration to generative unfolding is the central commitment of the GL architecture. It distinguishes GL from interactive proof assistants and from provers based on language models: both work toward goals that a human states.

The paper is organized as follows. \Cref{sec:workflow} describes the workflow of a run and \cref{sec:architecture} the architecture: MPL, conjecture generation, the counterexample filter, compilation, the processor and the compressor. \Cref{sec:mechanisms} describes the mechanisms of the prover: scopes, equivalence classes, proof methods, case splits, induction and the admission of new objects. \Cref{sec:mpu} outlines the hardware architecture and \cref{sec:output} the verifiable output. \Cref{sec:case_study} reports the results in elementary number theory, and \cref{sec:calculation} shows that GL also calculates. \Cref{sec:discovery} describes the current stage of the work, \cref{sec:related} compares GL with other discovery systems and with provers, and \cref{sec:agents} describes how AI agents operate GL and use it as a tool. \Cref{sec:outlook} gives the outlook and \cref{sec:conclusion} the conclusion.

\section{The GL Workflow}
\label{sec:workflow}

GL turns a set of definitions into a set of proved theorems through a deterministic pipeline. The main input is a set of clear and consistent definitions, written by a person or by an AI agent (\cref{sec:agents}). As in software development, an error in the definitions shows up as a compilation failure or as a contradiction during the run, and both help to correct the definitions.

A run consists of ordered batches. Each batch covers one theory layer, such as Peano arithmetic or Gauss's summation formula, and the theorems it proves are available to all later batches. Separate batches keep the search space of each layer small: a single search over all definitions would grow combinatorially with their number. A configuration file controls each batch. It defines the search space of the conjecturer, that is, which definitions, operators and structural templates may form conjectures, and it sets the depth of the prover, that is, how far the inference engine follows consequences before it stops. The configuration limits the search; it adds no axiom and no proof step.

\begin{figure}[htbp!]
    \centering
    \includegraphics[width=\textwidth]{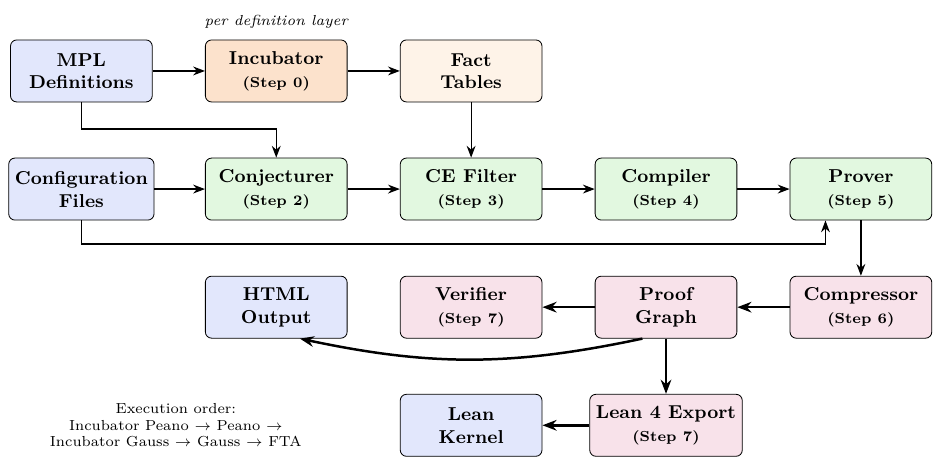}
    \caption{The GL pipeline. The MPL definitions feed both the Incubator (step~0) and the main pipeline, and the fact tables of the Incubator feed the CE filter. The configuration files of a batch limit the conjecturer and the depth of the prover. The proof graph is checked by the verifier, written as HTML and exported to Lean~4, where the Lean kernel checks it. A run executes the pipeline in ordered batches.}
    \label{fig:pipeline_flow}
\end{figure}

\begin{enumerate}
    \setcounter{enumi}{-1}
    \item \textbf{Incubation.} Before a theory layer is explored, the Incubator derives ground facts from its definitions, such as small addition and multiplication tables. It uses GL's own prover, with proofs by contradiction, to prove or disprove each candidate fact (\cref{sec:ce_filter}). These facts feed the counterexample filter (step~3), so the pipeline needs no input beyond the definitions. Each layer is incubated once, and established facts are reused.

    \item \textbf{Definition input.} A user or an AI agent writes the axioms and definitions in MPL, a language for a constrained subset of second-order predicate logic (\cref{sec:architecture}). Every later stage depends on the quality of this input.

    \item \textbf{Conjecture generation.} From the definitions, GL generates conjectures combinatorially. It weaves building blocks into a regular implication chain of the form $A \implies (B \implies (C \implies D))$, keeps one normalized representative of equivalent forms and discards malformed or redundant ones. This step is purely syntactic; it does not judge the value of a conjecture.

    \item \textbf{Counterexample (CE) filtering.} For each conjecture, GL's prover derives consequences of the conjecture together with the ground facts. A contradiction refutes the conjecture, and it is removed. A conjecture that survives is not proved; it is only consistent with the facts tested.

    \item \textbf{Compilation and disintegration.} The surviving conjectures are compiled onto a grid of Logic Blocks (LBs). Each premise of a conjecture gets its own LB, and the LBs of one conjecture form a chain; conjectures with common premises share LBs. Within each LB, the definitions are disintegrated into their logical parts, which are the starting material of the proof search.

    \item \textbf{Iterative proof execution.} GL tries to prove all conjectures at once, in iterations. An implication is stored as a hash-table entry: its premises form the key and its conclusion the value. In each iteration, every active LB combines its known statements into hash requests; a request that finds an entry fires the rule and yields a new statement together with its justification. The LBs work independently within an iteration and exchange statements only between iterations, so the search is parallel by construction (\cref{sec:architecture,sec:mpu}).

    \item \textbf{Post-proof processing.} The compressor removes redundancy: it keeps a minimal set of essential theorems from which the other proved theorems follow and passes them, with their proofs, to the output stage (\cref{sec:compressor}).

    \item \textbf{Verifiable output.} GL writes every proof as HTML pages in which each step links to its justification: a definition, an earlier step or a proved theorem. An independent verifier reads the processed proof graph and checks every step with a validator for its proof tag. Proofs can also be exported to Lean~4 and checked by the Lean kernel (\cref{sec:output}).
\end{enumerate}

GL is designed for hardware with many LBs working in parallel (\cref{sec:mpu}). On a single workstation, the configuration files make up for the smaller compute: the less compute is available, the more the configuration restricts the conjecturer and limits the depth of the prover. With more compute, the restrictions can be relaxed. At the other end of this range, the user supplies all needed lemmas as conjectures together with the definitions, and GL finds the order in which they follow from each other, and their proofs. In this way, from the Peano axioms, further definitions and a supplied list of conjectures, GL assembles a proof of the Fundamental Theorem of Arithmetic (FTA).

\section{System Architecture}
\label{sec:architecture}

The workflow runs on an architecture of many simple, independent processing elements, the Logic Blocks (LBs), which together perform the reasoning. The architecture can be realized in hardware (\cref{sec:mpu}) or in software on multi-core computers; the current implementation is software.

The architecture is not fixed. It is extended and reworked as GL is applied to more demanding mathematics; most mechanisms of \cref{sec:mechanisms} were added or changed during the work on the FTA. This section and the next therefore describe principles. Details are expected to change; the description corresponds to the release given in \cref{sec:availability}.

\subsection{Architecture Overview}

The core of GL is a symbolic inference engine that treats logical inference as a memory access. An implication is stored as an entry of a hash table. For an implication such as $(A \land B) \implies C$:
\begin{itemize}
    \item the premise, $A \land B$, is the \textbf{hash key};
    \item the conclusion, $C$, is the \textbf{hash value}.
\end{itemize}

Reasoning thus becomes the formulation of hash requests from known statements: a request that finds an entry returns a new, valid statement. For example, the definition of addition (\cref{lst:natural_numbers}) contains the rule
\begin{equation}
    b \in \mathbb{N} \;\wedge\; s(b) = c \;\wedge\; a + b = d \;\wedge\; s(d) = e \;\implies\; a + c = e,
\end{equation}
that is, $a + s(b) = s(a + b)$. Before the rule is stored, its premises are sorted by the name of their MPL predicate, \texttt{in} for $b \in \mathbb{N}$, \texttt{in2} for the two successor statements and \texttt{in3} for the sum, and its variables are renamed in the order of their first appearance. The key thus keeps only the pattern in which the variables repeat:
\begin{equation}
    \text{key: } x_1 \in \mathbb{N} \;\wedge\; s(x_1) = x_2 \;\wedge\; s(x_3) = x_4 \;\wedge\; x_5 + x_1 = x_3, \qquad
    \text{value: } x_5 + x_2 = x_4.
\end{equation}
Premises with the same predicate, here the two successor statements, are stored in each of their orders. An LB that knows the statements $n \in \mathbb{N}$, $s(n) = m$, $s(p) = q$ and $k + n = p$ combines them into a request, sorts them and renames their variables in the same way. The request has the same key, so the lookup finds the entry. The renaming gives $x_1 = n$, $x_2 = m$, $x_3 = p$, $x_4 = q$ and $x_5 = k$, and the value becomes the new statement $k + m = q$. Statements whose variables repeat in another pattern, such as $n \in \mathbb{N}$, $s(n) = m$, $s(p) = q$ and $k + r = p$, give another key and do not find this entry. A rule thus fires with one lookup, but an LB forms many requests in each iteration, and the work lies in forming them.

Retrieval by key is at the core of other inference engines as well. Saturation provers keep their clauses in index structures, such as discrimination trees and substitution trees, which return the stored terms that unify with a query term or are instances or generalizations of it \cite{sekar2001term}. Production systems match the patterns of all rules against a working memory with the Rete network, which keeps the state of partial matches between cycles and updates it only from changed elements \cite{forgy1982rete}. In GL, the key is the premise of an implication, and firing a rule is one lookup of that key. Keys and values are symbolic expressions. Internally, GL represents their symbols as integers and keeps the data of its proof search in statically allocated memory, as a hardware implementation would (\cref{sec:mpu}).

The engine runs on a grid of LBs. Each LB has its own memory and performs a burst of hash requests in each iteration. The LBs are connected along the premise chains of the conjectures (\cref{sec:compilation}): an LB reads the statements derived by the LBs above it and makes its own results available to the LBs below it. Within an iteration, the LBs do not communicate, so they can work in parallel.

\begin{figure}[htbp!]
    \centering
    \includegraphics[width=0.8\textwidth]{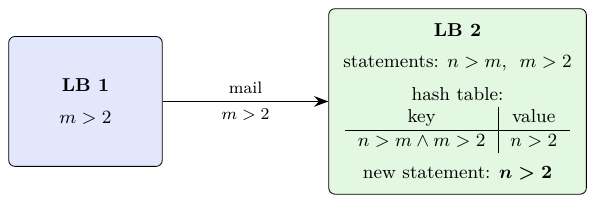}
    \caption{Two connected LBs. LB~2 receives the statement $m > 2$ from LB~1 by mail. Its hash table holds the key $n > m \land m > 2$ with the value $n > 2$, so a request from its known statements derives the new statement $n > 2$.}
    \label{fig:gl_grid_example}
\end{figure}

Consider the following example. Full implication:
\[
(n > m \land m > 2) \implies (n > 2)
\]
Premise:
\[
n > m \land m > 2
\]
Conclusion:
\[
n > 2
\]
\Cref{fig:gl_grid_example} shows in a simple way how two LBs can be connected.

\subsection{MPL}
\label{sec:mpl}

GL needs a formal, machine-readable language for definitions and axioms. MPL is based on a constrained subset of second-order predicate logic. It is expressive enough for the structures used in this paper and simple enough for algorithmic analysis and compilation; it belongs to the established framework of higher-order logic \cite{andrews2002logic}.

\subsubsection{LEs and Ports}
The basic unit of MPL is the logical entity (LE), a predicate with a Boolean value. An LE can be seen as a template function in a language such as C++: it receives its arguments through ports. For example, the LE for $x > 2$ has three ports: one for $x$, one for the constant $2$ and one for the relation $>$.

Each port has a \textbf{definition set}, which acts as its type. For $x > 2$, the ports for $x$ and $2$ may have the definition set $\mathbb{N}$, and the port for $>$ the definition set $\mathcal{P}(\mathbb{N} \times \mathbb{N})$, the set of binary relations on $\mathbb{N}$. The types allow strict checks at compile time.

\subsubsection{Core Operators}
New LEs are built from existing ones with three core operators, which suffice to express standard logic.

\begin{description}
    \item[Negation (\texttt{!})] \texttt{!} inverts the Boolean value of one LE. For example, \texttt{!(in[x,A])} means $x \notin A$. Ports and definition sets stay unchanged.

    \item[Conjunction (\texttt{\&})] \texttt{(\&AB)} combines two LEs $A$ and $B$ into one LE that is true if and only if both are true. Ports of $A$ and $B$ are merged by giving them the same name. In \texttt{(\&(in2[n,m,>])(in2[m,q,<]))}, the shared port \texttt{m} unifies the corresponding definition sets and links the two LEs.

    \item[Implication (\texttt{>})] \texttt{(>[vars](A)(B))} with an empty variable list is the implication $A \implies B$, which is equivalent to $\neg A \lor B$. With a non-empty list, as in \texttt{(>[n](A)(B))}, it is the universal quantifier: for every value of $n$ in the definition set of its port, $A(n) \implies B(n)$.
\end{description}

The existential quantifier is not a primitive: $\exists x: P(x)$ is written as $\neg(\forall x: \neg P(x))$. In the same way, a disjunction $A \lor B$ is written as $\neg(\neg A \land \neg B)$.

\subsubsection{Type System and Recursion Constraint}
The compiler enforces the types when ports are merged. One rule prevents infinite recursive definitions: the definition set of a port cannot be unified with a set that contains it. For example, a port of type $X$ cannot be merged with a port of type $\mathcal{P}(X \times Y)$, since this would make $X$ contain a structure defined by itself. The restriction, similar to type safety in programming languages, prevents paradoxes and keeps all definitions well-founded, although it limits the direct expression of concepts such as ordinal numbers.

\subsubsection{Unique Naming of Reusable LEs}
A reusable concept is defined once and given a name. \Cref{lst:fxy} shows the definition of a function $f: X \to Y$ in MPL. The LE \texttt{fXY} states that $f$ relates elements of $X$ to elements of $Y$, that every $x \in X$ has an image, and that this image is unique.

\begin{lstlisting}[language=MPL, caption={The definition of a function $f: X \to Y$ in MPL.}, label={lst:fxy}]
(fXY[f,X,Y]) :=
(&
    (>[x,y]
        (in2[x,y,f])
        (&
            (in[x,X])
            (in[y,Y])
        )
    )
    (&
        (>[x]
            (in[x,X])
            !(>[y]
                (in[y,Y])
                !(in2[x,y,f])
            )
        )
        (>[x]
            (in[x,X])
            (>[y1,y2]
                (&
                    (in2[x,y1,f])
                    (in2[x,y2,f])
                )
                (=[y1,y2])
            )
        )
    )
)
\end{lstlisting}

The compiler also names the inner parts of a definition, such as an existence statement, so that equal parts of different definitions share one name. \Cref{lst:natural_numbers} shows the definition of the natural numbers, with zero, the successor function, addition and multiplication, as it is supplied to GL.

\begin{lstlisting}[language=MPL, caption={The definition of the natural numbers in MPL.}, label={lst:natural_numbers}]
(NaturalNumbers[N,i0,s,+,*]) :=
(&
    (&
        (in[i0,N])
        (&

            (fXY[s,N,N])
            (&
                (>[n]
                    (in[n,N])
                    !(in2[n,i0,s])
                )
                (>[m]
                    (in[m,N])
                    (>[n1,n2]
                        (&
                            (in2[n1,m,s])
                            (in2[n2,m,s])
                        )
                        (=[n1,n2])
                    )
                )
            )
        )
    )
    (&
        (&
            (fXYZ[+,N,N,N])
            (&
                (&
                    (>[a]
                        (in[a,N])
                        (>[b]
                            (in3[a,i0,b,+])
                            (=[a,b])
                        )
                    )
                    (>[a,b]
                        (&(=[a,b])(&(in[a,N])(in[b,N])))
                        (in3[a,i0,b,+])
                    )
                )
                (>[b]
                    (in[b,N])
                    (>[a,c,d]
                        (&
                            (in2[b,c,s])
                            (in3[a,b,d,+])
                        )
                        (&
                            (>[e]
                                (in3[a,c,e,+])
                                (in2[d,e,s])
                            )
                            (>[e]
                                (in2[d,e,s])
                                (in3[a,c,e,+])
                            )
                        )
                    )
                )
            )
        )
        (&
            (fXYZ[*,N,N,N])
            (&
                (&
                    (>[a]
                        (in[a,N])
                        (>[b]
                            (in3[a,i0,b,*])
                            (=[b,i0])
                        )
                    )
                    (>[a,b]
                        (&(=[b,i0])(&(in[a,N])(in[b,N])))
                        (in3[a,i0,b,*])
                    )
                )
                (>[b]
                    (in[b,N])
                    (>[a,c,d]
                        (&
                            (in2[b,c,s])
                            (in3[a,b,d,*])
                        )
                        (&
                            (>[e]
                                (in3[d,a,e,+])
                                (in3[a,c,e,*])
                            )
                            (>[e]
                                (in3[a,c,e,*])
                                (in3[d,a,e,+])
                            )
                        )
                    )
                )
            )
        )
    )
)
\end{lstlisting}

\paragraph{The anchor.} Every conjecture of a batch begins with the same premise, the anchor of the batch: one atomic statement that fixes the structure in which the batch works. For Peano arithmetic it is $\texttt{AnchorPeano}[\mathbb{N}, 0, s, +, \cdot, 1]$, which states $\texttt{NaturalNumbers}[\mathbb{N}, 0, s, +, \cdot]$ (\cref{lst:natural_numbers}) and $s(0) = 1$. The anchors of the Gauss and FTA batches add $s(1) = 2$ and the identity function on $\mathbb{N}$, and the anchor of the Incubator names the numerals $0, \ldots, 8$. The arguments of the anchor are the constants of the batch; GL does not induct on them (\cref{sec:induction}).

\subsection{Creation of Conjectures}
\label{sec:conjectures}

After the definitions are written in MPL, GL creates conjectures automatically. The process is combinatorial, and several layers of control keep the number of conjectures manageable. The aim is a rich set of candidate theorems without a flood of statements that are syntactically valid but useless. Theory exploration systems also generate candidate conjectures from the given functions: QuickSpec \cite{claessen2010quickspec} filters them by testing, and IsaCoSy \cite{johansson2011conjecture} filters them with counterexample finders and proves the survivors (\cref{sec:related}); GL generates implication chains over its definitions.

The main control is a regular theorem structure. Many theorems can be written as, or easily turned into, a linear chain of implications such as $A \implies (B \implies (C \implies D))$. GL generates conjectures of this form by weaving: it starts with a base LE and connects further building blocks to it with the implication operator. Further controls act during the weaving:

\begin{enumerate}
    \item \textbf{Normalization.} After each new building block, all port names are renamed by a deterministic, canonical scheme. Among all orders of the building blocks in the implication chain, GL keeps the lexicographically smallest. Equivalent statements such as $A \implies (B \implies C)$ and $B \implies (A \implies C)$ therefore yield one conjecture.

    \item \textbf{Type-based connection filtering.} The type system restricts which ports may be connected. In arithmetic, for example, the configuration can permit connections only between ports with identical definition sets.

    \item \textbf{Topological filtering.} Rules exclude conjectures with certain structures, for example a circular dependency in which the output of an LE is indirectly used as its own input.
\end{enumerate}

With these controls, the conjecturer produces 1,001 conjectures for Peano arithmetic.

Three further ideas prepare case splits (\cref{sec:branching}). First, a conjecture with a negated premise, $\neg A \implies B$, has the contrapositive $\neg B \implies A$; before the proof search, the prover adds it to the conjectures that passed the CE filter and proves it like any other. Second, the configuration marks expressions that may be negated and expressions that may form an existence statement; from each such pair, the conjecturer generates two implications that together state a disjunction, for example that every natural number is zero or has a predecessor. Third, a proved implication $\neg A \implies B$ is equivalent to the disjunction $A \lor B$; GL registers this disjunction as an OR theorem, which later proofs use to split cases. The main development of the conjecturer in this direction follows in the current campaign (\cref{sec:discovery}).

\subsection{CE Filter}
\label{sec:ce_filter}

The counterexample (CE) filter removes conjectures that contradict known ground facts. It reuses the LB principle with a flat distribution: each conjecture gets its own LB, which receives the conjecture and a finite set of ground facts, such as small addition and multiplication tables. Within its LB, the filter weaves facts toward the premises of the conjecture with a peek-and-prune strategy: at each step it looks ahead just enough to decide whether to continue a branch or to cut it. When a derived fact contradicts a ground fact, the conjecture is rejected and its LB stops.

The filter is incomplete as a decision procedure: a conjecture that survives is not proved, only compatible with the facts tested. Its cost does not dominate the run time or the memory, and it removes most false conjectures before the proof search. Proof assistants filter conjectures in a similar way: Nitpick searches for a finite model of the negated conjecture with a SAT-based model finder \cite{blanchette2010nitpick}, and Quickcheck evaluates the conjecture on concrete values \cite{bulwahn2012new}. GL's filter uses its own prover and the ground facts of the Incubator.

\paragraph{Incubator.}
The ground facts come from the Incubator, which uses GL's own prover. Its conjecturer proposes candidates over the named numerals of its anchor, and the prover proves or refutes them. A false candidate, such as $3 + 4 = 8$, is assumed, and a contradiction proves its negation, $3 + 4 \neq 8$. The value of an operator is proved as an implication, such as $3 + 4 = x \implies x = 7$, and the fact $3 + 4 = 7$ is obtained from it by back-reformulation (\cref{sec:calculation}). The inequalities and the facts about the order are thus proved theorems, and the operator facts come from proved implications. The Incubator is guided by its own configuration files, one per batch. In richer domains, the ground facts are no longer table entries but statements that need a structured derivation, and the Incubator becomes more important.

\subsection{Compilation}
\label{sec:compilation}

The compilation translates each conjecture into a configuration of the LB grid. It has three parts: mapping, disintegration and resource sharing.

\subsubsection{Mapping to the Logic Grid}
\begin{figure}[htbp!]
    \centering
    \includegraphics[width=0.85\textwidth]{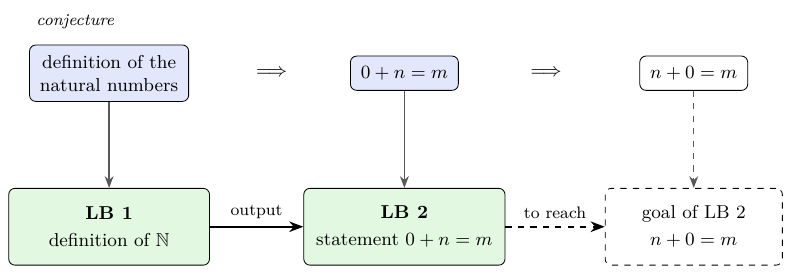}
    \caption{Mapping the conjecture $0 + n = m \implies n + 0 = m$ onto a chain of LBs. LB~1 holds the definition of the natural numbers, LB~2 the statement $0 + n = m$, and the conclusion $n + 0 = m$ is the goal of LB~2.}
    \label{fig:compilation_grid}
\end{figure}

Each conjecture is mapped onto a chain of LBs: the building blocks of its premise are assigned to a sequence of LBs. For example, the theorem $\forall n, m \in \mathbb{N}: 0 + n = m \implies n + 0 = m$ has a premise of two LEs, the definition of the natural numbers and the statement $0 + n = m$, and the conclusion $n + 0 = m$. The two LEs are mapped to two LBs connected in a directed path, where the output of the first is available to the second (\cref{fig:compilation_grid}). The conclusion is not mapped to an LB; it is the goal that the last LB of the chain has to reach.

\subsubsection{Disintegration}
Within its LB, an LE is disintegrated into its parts. Each sub-expression, such as a named function or a built-in operator, is extracted and stored in the local memory of the LB. An implication inside a definition is not disintegrated further; it is stored in the hash table of the LB and used as a rule. An existence statement such as $\exists y: P(x, y)$ is broken down by existential instantiation: a new, locally named port, a witness, is created for the bound variable, so that the LB can reason about a concrete instance. The witness plays the role that a Skolem constant plays in resolution provers \cite{nonnengart2001computing}. These parts are the material that the LB recombines during execution.

\subsubsection{Resource Sharing via Tree Structuring}
\begin{figure}[htbp!]
    \centering
    \includegraphics[width=0.8\textwidth]{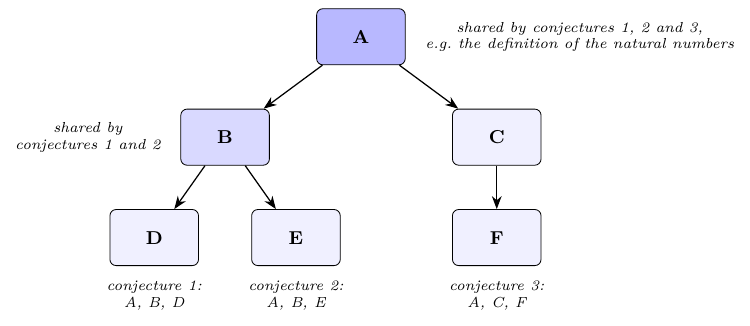}
    \caption{Shared tree structure for three conjectures. Conjectures with a common beginning of their premise chains share its LBs: all three share LB~A, for example the definition of the natural numbers, and conjectures 1 and 2 also share LB~B. The chains branch where the premises differ.}
    \label{fig:theorem_tree}
\end{figure}

Conjectures with a common prefix of their implication chain, for example the definition of the natural numbers, share the LBs of that prefix. The shared LBs form the root of a tree whose branches carry the different parts of the conjectures (\cref{fig:theorem_tree}). Each common part is therefore processed once.

\subsection{Processor}
\label{sec:processor}

The processor executes the compiled conjectures on the LB grid; this is where the proof search takes place.

\subsubsection{The LB Step}
A run consists of iterations. In each iteration, every active LB performs one step of three phases:

\begin{enumerate}
    \item \textbf{Mail absorption.} The LB takes in the statements and rules that the LBs above it derived since its last step. Mail is implemented as reads: each LB keeps a log of its results, and an LB reads the logs of its ancestors.

    \item \textbf{Hash burst.} The LB combines its known statements into hash requests and fires every rule whose key a request finds.

    \item \textbf{Inner mail absorption.} The statements produced by the hash burst reach the LB as inner mail. The LB absorbs them like incoming mail: it registers them, brings them into canonical form with respect to the known equalities (\cref{sec:equivalence}) and checks them against its goals.
\end{enumerate}

All LBs complete a phase before the next phase starts. The result of a step therefore does not depend on the order in which parallel workers process the LBs. The iterations resemble the stages of bottom-up evaluation in Datalog, which applies the rules until no new fact appears \cite{abiteboul1995foundations}. Unlike Datalog rules, GL's rules can introduce new objects through existence statements; \cref{sec:admission} describes how GL limits them. A theorem proved in an LB is registered globally and made available to all LBs as a rule.

\subsubsection{The Hash Burst}
The hash burst works on three kinds of data: statements derived in the LB, statements received from its ancestors, and the rules in its hash table, which come from definitions and proved theorems. The LB grows combinations of statements step by step, in a variant of the peek-and-prune family of algorithms: after each extension, it checks whether the combination is the beginning of some key in its hash table, and it abandons the branch if it is not. Every request must involve a statement or a rule that the LB owns. This distributes the work: an inference is computed in the LB that owns its ingredients, not in every LB that can see them.

\subsubsection{Deterministic Logical Transformer}
The hash engine that applies proved theorems, that is, performs algebraic manipulation, has a second task: the deterministic transformation of existing LEs into new ones. We call this capability the \textbf{Deterministic Logical Transformer}. (The term ``transformer'' is used in its classical sense, a mechanism that converts one structure into another; it is not related to the neural Transformer architecture.)

The transformation consists of two complementary operations: \textit{disintegration} and \textit{integration}. Disintegration decomposes a compound LE into its parts; integration assembles available parts into new, valid compound LEs. Both are governed by the same artifact, the GL binary, which is the main output of the compilation. The binary encodes exactly how a compound structure may be taken apart and how its parts may be recombined. The compiled representation has no direction: the same map can be traversed both ways, so disintegration and integration are two directions of one deterministic process.

\begin{figure}[ht]
    \centering
    \includegraphics[width=\linewidth]{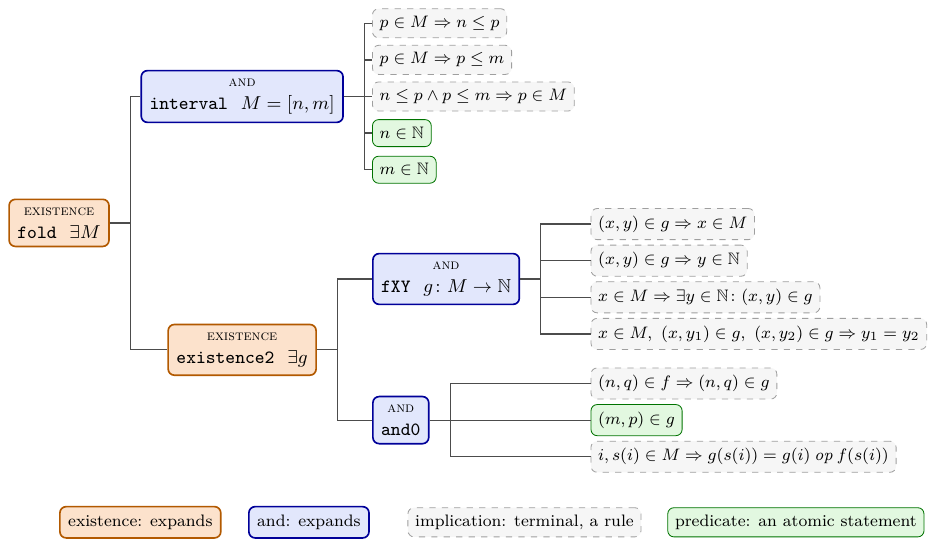}
    \caption{Compiled structure of the \texttt{fold} definition. Existence and and-nodes expand further; implication nodes are terminal; predicate leaves are atomic statements. The Deterministic Logical Transformer traverses the same map in both directions.}
    \label{fig:fold_graph}
\end{figure}

\Cref{fig:fold_graph} illustrates this structure for the \texttt{fold} definition. The root node is of type \textit{existence} and expands into two children: \texttt{interval} (type \textit{and}) and \texttt{existence2} (type \textit{existence}). Each of these expands further into its elements. Nodes of type \textit{implication} are terminal: they are not expanded but stored as inference rules in the local hash table of the corresponding LB. Predicate leaves are atomic statements that serve directly as facts or hash keys during execution. The same graph that guides disintegration, which breaks \texttt{fold} down to its atomic parts, also guides integration, which assembles a valid \texttt{fold} instance from available parts by traversing the map in reverse.

The two directions use the node types differently. During disintegration, \textit{and} and \textit{existence} nodes supply predicates: their elements are extracted and passed to the hash engine as known facts. \textit{Implication} nodes are loaded into the hash table as key-value pairs and serve as rules. During integration, the roles change. Implication nodes become sub-theorems that must be proved before the compound LE can be assembled. \textit{And} and \textit{existence} nodes are turned into implications: the question whether a compound LE can be constructed becomes a conditional statement, which the hash engine can establish when the necessary parts are present. Disintegration and integration thus use the same machinery in opposite directions: one dissolves structure into material, the other builds structure from proofs.

\subsection{Compressor}
\label{sec:compressor}

A run proves many theorems that follow from others. The compressor reduces the proved theorems to an essential core, a small set from which all the others can be derived. Only this core is passed to later batches as proved knowledge, which keeps the rule memory of the later batches small. The compressor changes no proof; it selects theorems. Saturation provers remove redundant clauses in a similar spirit, those that follow from smaller clauses of the set \cite{bachmair2001resolution}, and QuickSpec drops a candidate equation that follows from the equations it has already found \cite{smallbone2017quick}. Both remove redundancy during the search; GL's compressor works after the proofs, on the proved theorems as a whole.

It works in two stages. First, it creates one LB for each proved theorem, with all proved theorems as rules, that theorem included, the premises of that theorem as statements and its conclusion as the goal, and it records the derivations of the goal that this LB finds, with the theorems they use. Second, it removes theorems greedily, the least used first. A removal is kept only when every proved theorem, the removed ones included, can still be derived from its premises and the remaining theorems by the recorded derivations; a derivation that uses a removed theorem, the tested theorem itself included, no longer counts. Otherwise the theorem is essential. Two theorems that can be derived only from each other therefore cannot both be removed. With respect to the recorded derivations, no theorem of the result can be removed without losing the derivation of another, although the result is not necessarily the smallest such set.

In the FTA discovery (\cref{sec:discovery}), the compressor keeps the growing number of proved theorems manageable: each step continues from the logical core of the previous one.

\section{Mechanisms of the Prover}
\label{sec:mechanisms}

This section describes three mechanisms of the prover: scopes, which separate what holds under different assumptions; equivalence classes, which handle equalities; and case splits over disjunctions. It also describes the proof methods that use them, and induction in detail. The last subsection describes when GL creates new objects.

\subsection{Scopes}
\label{sec:scopes}

Every statement in an LB holds in a scope. The main scope contains what follows from the premises of the conjecture. An inner scope contains what follows under one additional assumption, for example:
\begin{itemize}
    \item the premise of an implication that the LB tries to prove (a subproof);
    \item one case of a disjunction (\cref{sec:branching});
    \item the negation of a statement that the LB tries to prove (a proof by contradiction).
\end{itemize}

The scopes of an LB form a tree. An inner scope sees every statement of its ancestors; an outer scope does not see the statements of its inner scopes. When the goal of an inner scope is reached, the result moves to its parent scope without the assumption, and the inner scope with everything derived in it is removed. A scope whose assumption turns out to be false is removed as well.

For example, to prove $x \in A \implies x \in B$ in an LB, GL opens an inner scope that assumes $x \in A$. The rules of the LB fire in this scope on its own statements and on the statements of the main scope. When $x \in B$ is derived in the inner scope, the implication $x \in A \implies x \in B$ holds in the main scope, and the inner scope is removed.

Inner scopes play the role of the subproofs of natural deduction \cite{gentzen1935untersuchungen,fitch1952symbolic}. In Fitch's form, a hypothesis opens a subordinate proof, which may use the formulas of the enclosing proofs. Nothing derived in it may be used outside, and when $B$ is derived in a subordinate proof with the hypothesis $A$, the enclosing proof gains $A \implies B$ \cite{fitch1952symbolic}. GL opens such subproofs for the tasks listed above, and the rules of the LB fire in them as in the main scope.

Each statement also records which premises of the conjecture it depends on. A conjecture counts as proved only when its proof uses all its premises. A proof that uses fewer premises shows a stronger statement; GL records it, but does not register the conjecture as a theorem. The stronger statement belongs in the knowledge base in its own form, without the unused premises; registering the weaker form would add a theorem with a premise too many. This record resembles the labels of an assumption-based truth maintenance system, which marks every derived datum with the sets of assumptions under which it holds and so works in many contexts at once, without backtracking \cite{dekleer1986assumption}. In GL, the record covers the premises of the conjecture, and the scopes hold the further assumptions.

\subsection{Equivalence Classes}
\label{sec:equivalence}

Equalities need special treatment. A rule for $=$ would produce, for every equality, copies of all statements with one term replaced by the other, and the number of statements would grow quickly. GL therefore collects equal terms into equivalence classes, one set of classes per scope, and chooses one member of each class as its representative. A statement is stored only in its canonical form, in which every term is replaced by its representative, and the rewriting is recorded as a proof step that cites the equalities used.

For example, if an LB knows $a = b$ and $b = c$, the terms $a$, $b$ and $c$ form one class with the representative $a$. A statement derived about $c$, say $c \in \mathbb{N}$, is stored as $a \in \mathbb{N}$, together with the proof step that rewrites $c$ to $a$. Statements that differ only by equal terms thus coincide.

GL does not create all variants of a statement for all members of a class; it creates a variant only when a rule needs it as a premise. An inner scope inherits the classes of its ancestors and may join further classes under its own assumption; such joins hold only in the inner scope.

Classes of equal terms are a common tool of automated reasoning. The union-find structure keeps a partition into disjoint sets, with one operation that returns the name of the set of an element and one that merges two sets \cite{tarjan1975efficiency}. Congruence closure adds the rule that two applications of a function are equal when their arguments are equal; with it, the satisfiability of a conjunction of equations and negated equations can be decided \cite{nelson1980fast,downey1980variations}. An e-graph extends union-find to classes of expressions that are closed under this rule. In equality saturation, rewrite rules only add equalities to an e-graph, and the library egg alternates a phase that finds all matches of the rules with a phase that merges the matched classes; it restores the invariant of the e-graph at the boundaries between these phases, not after every operation \cite{willsey2021egg}. The LB step of GL has a similar order: the hash burst fires every rule whose key a request finds, and the next phase brings the new statements into canonical form (\cref{sec:processor}).

\subsection{Proof Methods}
\label{sec:proof_methods}

Most proofs are direct: the LB at the end of the premise chain derives the conclusion. Two further methods use the same engine.

\paragraph{Contradiction.} To prove a conjecture $A \implies (B \implies (C \implies D))$ by contradiction, GL maps it onto the chain $A \to B \to C$ and attaches the negated conclusion $\neg D$ as a fourth LB, which is primed to detect a contradiction. When this LB holds a statement and its negation, for example $P(x)$ and $\neg P(x)$, the conjecture is proved and registered as a theorem.

\paragraph{Induction.} For a conjecture about all natural numbers, GL also proves by induction (\cref{sec:induction}).

\subsection{Branching in the Prover}
\label{sec:branching}

Many proofs need a split into cases. GL represents a disjunction $E_1 \lor \dots \lor E_k$ as one compiled entity, an OR, whose disjuncts keep their own signs.

\paragraph{Sources.} An OR enters a proof from a definition that contains one, from an OR theorem (\cref{sec:conjectures}) or from a conjecture that states one. For example, the OR theorem that every natural number $n$ is $0$ or the successor of some $m$ is the basis of many case splits in arithmetic.

\paragraph{Case split.} A known OR does not split the proof by itself. GL opens a case split only in an LB with open proof work: when the OR is a premise of the conjecture or of an implication that the LB tries to prove, or when a rule needs one of its disjuncts as a premise. Each case gets its own scope below the scope in which the OR holds, with the disjunct as its assumption. A nested OR, whose disjuncts are again ORs, is flattened into one list of cases.

\paragraph{Order.} The cases are processed one at a time, in a fixed order: cases whose assumption is not an equality come first, and equalities with a constant of the anchor, such as $n = 0$, come last. The next case opens only when the current one is resolved. A case is resolved in one of two ways: its scope reaches the goal of the enclosing scope, or its assumption is refuted by statements known above it, and the case is retired.

\paragraph{Convergence.} When every case that is not retired has reached the goal, the goal holds in the scope of the OR. Its record of premises is the union of the premises used by all cases, including those used to retire cases. The case scopes are then removed.

\begin{figure}[htbp!]
    \centering
    \includegraphics[width=0.85\textwidth]{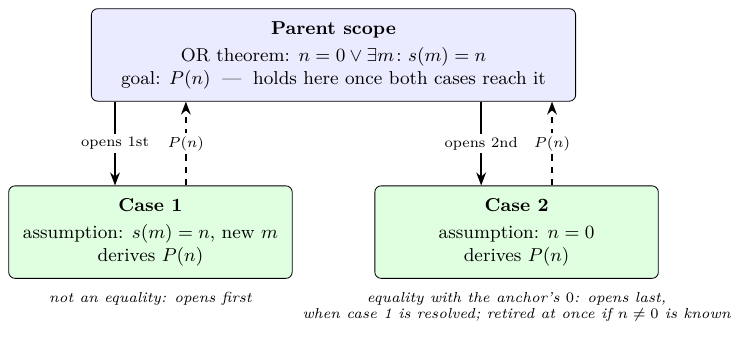}
    \caption{A case split in the scope tree of an LB, for the example in the text. The two case scopes open one after the other, each with its disjunct as its assumption. When both reach $P(n)$, the goal holds in the parent scope and the case scopes are removed.}
    \label{fig:case_split}
\end{figure}

For example (\cref{fig:case_split}), to prove a statement $P(n)$ with the OR theorem $n = 0 \lor \exists m\colon s(m) = n$, GL first opens a case scope that assumes $s(m) = n$ for a new $m$. When this case reaches $P(n)$, the second case scope opens and assumes $n = 0$. When this case reaches $P(n)$ as well, $P(n)$ holds in the parent scope. If $n \neq 0$ is already known, the case $n = 0$ is refuted and retired, and the first case alone decides.

\paragraph{Exclusion.} A known OR also acts without a case split: when all disjuncts but one are refuted, the remaining one holds, and when some of them are refuted, the OR of the remaining disjuncts holds.

\paragraph{Negations and goals.} The negation of an OR is the conjunction of the negated disjuncts and needs no case split. To prove an OR as a goal, GL assumes the negations of the other disjuncts and proves the remaining one.

\paragraph{Merging variants.} When a theorem is proved in variants, one for each disjunct of a proved OR theorem, GL merges the variants into one theorem without the case condition.

Every case split is recorded in the proof graph with its cases, and the verifier checks that the cases cover the OR and that each case reaches the goal or is refuted.

Case splits are a basic step of many provers. The procedure of Davis, Logemann and Loveland splits a set of clauses on a propositional variable $p$ into two smaller problems, one with $p$ true and one with $p$ false; it stores one of them while it works on the other, so the cases are processed one at a time \cite{davis1962machine}. In analytic tableaux, a disjunctive rule splits a branch of the tableau into alternative branches, and a formula is proved when every branch closes, that is, contains a contradiction \cite{smullyan1968first}. AVATAR splits clauses into parts without shared variables and lets a SAT solver choose one part of each split clause; the first-order prover works on the chosen parts, and when it refutes them, the SAT solver receives a new clause and chooses again \cite{voronkov2014avatar}. In GL, the cases are scopes, processed one at a time in a fixed order; a case ends when it reaches the goal or when its assumption is refuted, and a split is opened only when a premise or a rule needs it.

\subsection{Induction}
\label{sec:induction}

GL proves a conjecture about all natural numbers by induction. The induction principle is part of the prover, not of the definitions. We write the conjecture as
\begin{equation}
    \Phi(n, \vec{y}) \;=\; \bigl(A(n, \vec{y}) \implies C(n, \vec{y})\bigr),
\end{equation}
where $n$ is the induction variable, $\vec{y}$ are the other variables, $A$ is the chain of premises and $C$ is the conclusion.

\paragraph{Three parts.} An induction consists of three parts, each recorded as its own proof chapter:
\begin{align}
    \text{type:}\quad & A(n, \vec{y}) \;\vdash\; n \in \mathbb{N}, \\
    \text{zero:}\quad & A(n, \vec{y}),\; n = 0 \;\vdash\; C(n, \vec{y}), \\
    \text{step:}\quad & A(n, \vec{y}),\; s(\mathit{rec}) = n,\; H \;\vdash\; C(n, \vec{y}).
\end{align}
The type part is derived in the last LB of the conjecture's chain. It makes sure that the induction ranges over the natural numbers; without it, the conclusion would be claimed for objects that are not numbers. The zero part and the step part get their own LBs below that LB, with $n = 0$ and with $s(\mathit{rec}) = n$ as their additional premise; they receive the premises of the conjecture from the chain above them. $H$ is the induction hypothesis. The zero part starts only after the step part has succeeded. GL registers the theorem when all three parts are proved and together use all premises of the conjecture.

\paragraph{The step goes from $\mathit{rec}$ to $n$.} The textbook step derives the statement for $n + 1$ from the statement for $n$. GL keeps the conjecture's own variable $n$ in the goal and introduces its predecessor $\mathit{rec}$:
\begin{equation}
    \underbrace{s(\mathit{rec}) = n}_{n \text{ is a successor}}, \qquad
    \underbrace{H(\mathit{rec})}_{\text{hypothesis at } \mathit{rec}}
    \;\vdash\; C(n, \vec{y}).
\end{equation}
The step thus derives the conjecture for $n = s(\mathit{rec})$ from the hypothesis at $\mathit{rec}$, and the zero part proves it for $n = 0$. By the induction principle, the two parts together give the conjecture for every $n$. The goal and the premises of the step are those of the conjecture; only $s(\mathit{rec}) = n$ and the hypothesis are new.

\paragraph{The hypothesis.} The hypothesis $H$ is an auxiliary implication that GL builds from the conjecture and stores in the hash table of the step LB as a rule. It has one of three forms.

\emph{Weak.} The induction variable is replaced by $\mathit{rec}$. The other variables that GL could induct on (see below), $\bar{\vec{y}}$, keep the values they have in the goal; all remaining variables, $\vec{z}$, are universal:
\begin{equation}
    H_{\text{weak}} \;=\; \forall \vec{z}\colon\; A(\mathit{rec}, \bar{\vec{y}}, \vec{z}) \implies C(\mathit{rec}, \bar{\vec{y}}, \vec{z}).
\end{equation}
For the commutativity of addition with induction on $a$, the fixed variable is $b$, and $c$, the result of the addition, is universal:
\begin{equation}
    \Phi(a, b, c) = \bigl(a + b = c \implies b + a = c\bigr), \qquad
    H_{\text{weak}} = \forall c'\colon\; \mathit{rec} + b = c' \implies b + \mathit{rec} = c'.
\end{equation}

\emph{Stronger.} Some conjectures have twin premises: two premises with the same predicate that differ in the induction variable $k$ and a partner $l$. GL then inducts on $k$ only, for the statement over all values of the partner: the partner stays universal in the hypothesis. The hypothesis therefore holds for every value of $l$, and the step can apply it at the predecessor of $l$. The uniqueness part of the FTA needs this form. With $\mathit{pfs}(k, P, n)$ for ``$P$ is a nonincreasing sequence of primes with last index $k$ and product $n$'', the conjecture and its hypothesis are
\begin{align}
    \Phi &= \bigl(\mathit{pfs}(k, P, n) \wedge \mathit{pfs}(l, Q, n) \implies P = Q\bigr), \\
    H_{\text{stronger}} &= \forall P', n', l', Q'\colon\; \mathit{pfs}(\mathit{rec}, P', n') \wedge \mathit{pfs}(l', Q', n') \implies P' = Q'.
\end{align}
In the step, GL splits the partner into the cases $l = 0$ and $l = s(\mathit{rec}_2)$ and names its predecessor $\mathit{rec}_2$ in the same way as $\mathit{rec}$. The case $l = 0$ is refuted, and the hypothesis is applied to the two shortened sequences $R$ and $R'$ with their common product $m$:
\begin{equation}
    s(\mathit{rec}) = k, \quad s(\mathit{rec}_2) = l, \qquad
    \mathit{pfs}(\mathit{rec}, R, m) \wedge \mathit{pfs}(\mathit{rec}_2, R', m) \implies R = R'.
\end{equation}

\emph{Strong.} The hypothesis holds at every value below $n$. Only $n$ is fixed; every other variable is universal. The bound $m < n$ is written with atomic statements, not with the defined order:
\begin{equation}
    H_{\text{strong}} \;=\; \forall m, x, \vec{z}\colon\; m + x = n \;\wedge\; m \neq n \;\wedge\; A(m, \vec{z}) \implies C(m, \vec{z}).
\end{equation}
The step LB asks for values below $n$: when an existence statement provides a witness $m$ with $m + x = n$, the witness is admitted, and the hypothesis can fire at $m$. A form of B\'ezout's identity for coprime natural numbers, proved by induction on $b$, needs this form. It divides $a$ by $b$ and applies the hypothesis at the remainder, with $b$ in the place of $a$:
\begin{equation}
    a = q \cdot b + r, \quad r + x = b, \quad r \neq b \qquad \Longrightarrow \qquad \Phi(b, r).
\end{equation}
The weak form offers only the predecessor of $b$, with $a$ fixed.

A second hypothesis of the same kind, the multiplication catcher, bounds $m$ by a product instead of a sum:
\begin{equation}
    H_{\text{catch}} \;=\; \forall m, x, \vec{z}\colon\; x \cdot m = n \;\wedge\; m \neq n \;\wedge\; A(m, \vec{z}) \implies C(m, \vec{z}).
\end{equation}
In the step, $n = s(\mathit{rec})$ is not $0$, so $x \cdot m = n$ and $m \neq n$ imply $m < n$. The catcher acts through its demand $\square_1 \cdot \square_2 = n$, which admits a cofactor of $n$ as a value below $n$; the hypothesis that then fires at this value is $H_{\text{strong}}$. The existence part of the FTA uses it. For $n \geq 2$, the least divisor $p \geq 2$ of $n$ is prime and gives $p \cdot d = n$, and the catcher admits the cofactor $d$. For $d = 1$, the sequence $(p)$ is the factorization of $n$. For $d \geq 2$, the strong hypothesis, fired at $d$, gives a prime factorization $F$ of $d$. Every prime factor of $d$ divides $n$ and is therefore at least $p$, so $F$ with $p$ appended at the end is again nonincreasing and factorizes $n$. With sequences indexed from $0$, as in \cref{fig:proof_chapter_screenshot}:
\begin{equation}
\begin{aligned}
    & n \geq 2, \quad p = \min\{q \geq 2 : q \mid n\}, \quad p \cdot d = n, \quad d \neq n, \quad 2 \leq d, \\
    & \text{hence } F = (F_0, \ldots, F_k) \text{ factorizes } d, \text{ hence } (F_0, \ldots, F_k, p) \text{ factorizes } n.
\end{aligned}
\end{equation}

\paragraph{Forward eating.} The step LB uses the hypothesis along one line of reasoning. The premises of $H$ form a chain, and the LB fills them in order with statements it knows. Each filled premise is eaten, and the rest of the chain remains:
\begin{equation}
    \underbrace{B_1 \implies \bigl(B_2 \implies D\bigr)}_{H}
    \;\xrightarrow{\;\text{eat } B_1\;}\;
    B_2 \implies D
    \;\xrightarrow{\;\text{eat } B_2\;}\;
    D.
\end{equation}
When a premise is filled with a new witness, the eating continues in a scope of its own, an induction scope. Alternative witnesses for the same premise open sibling scopes, which exchange no statements. The scopes thus form a tree, and no request combines statements of two of its paths. The premises are eaten in one order on each path, and an eaten premise is not filled again. The first time the hypothesis fires in the tree, the rest of the tree is removed, and the hypothesis opens no further induction scope in it. An induction scope carries no assumption, so a goal reached in it holds in the scope above. In this way a tree uses the hypothesis once, along one line of reasoning, as in a proof written by hand. After this firing, the hypothesis can still fire on statements that the remaining path already knows, but it admits no new value. A strong hypothesis never fires on a statement that is one of its own conclusions or derives from one. A proof that needs the hypothesis at two values is therefore found only when the second value is already known on the same path. The work of the step is a sum over the paths of the tree; with several witnesses for each premise, the number of paths can be the product of their numbers. A proof that needs statements of two sibling scopes before either reaches the goal is not found. Rippling controls the search in the induction step by the syntactic similarity between the induction conclusion and the hypotheses, which it marks with annotations on the formulas \cite{bundy2005rippling}. In GL, the hypothesis is a rule of the step LB, and forward eating applies it along one line of reasoning.

\paragraph{Choice of the induction variable.} GL inducts only on variables that range over the natural numbers and appear in the conjecture only at inputs of operators, relations or properties. A variable that also appears at an output, or at a place that is neither an input nor an output, is not a candidate, and neither are the arguments of the anchor ($\mathbb{N}$, $0$, $s$, $+$, $\cdot$). For example,
\begin{equation}
    \underbrace{a}_{\text{input}} + \underbrace{b}_{\text{input}} = \underbrace{c}_{\text{output}}
    \;\Longrightarrow\; \text{candidates } a, b;
    \qquad
    \mathit{pfs}(\underbrace{k}_{\text{input}}, \underbrace{P}_{\text{input}}, \underbrace{n}_{\text{output}})
    \;\Longrightarrow\; \text{candidate } k.
\end{equation}
The input $P$ is a sequence, not a number. For every candidate and every enabled form, GL prepares one induction, and all of them run in parallel. GL prepares the strong form only for some conjectures: every premise that contains the candidate is a property of it, a relation with another candidate or an operator that defines a set from it, and no statement of the conjecture applies an operator whose value is a number, such as $a + b = c$. If two candidates qualify, the strong form is prepared only when the premises are symmetric in them; with three or more, it is not prepared. The first step part that reaches its goal wins, provided its proof and the type part together use all premises of the conjecture: its zero part starts, and the other inductions and the proofs by contradiction of the conjecture are switched off. The order is decided at the end of each iteration: of the step parts that reached their goals in that iteration, the first in a fixed order of their LBs wins, so the choice does not depend on timing. If the zero part of the winner then fails, the switched-off routes are not resumed: the conjecture can still be proved directly, and otherwise it stays open in this run. With twin premises, only the variable of the first twin is a candidate. The Boyer--Moore prover chooses one induction scheme for a conjecture: each recursive call in the conjecture suggests a scheme, the suggested schemes are merged, and heuristics pick the best \cite{boyer1979computational}. GL does not choose one; it tries all candidates in parallel. Induction has also been added to automated provers: to SMT solving in CVC4 \cite{reynolds2015induction}, to superposition with structural induction \cite{cruanes2017superposition} and to saturation-based proof search in Vampire \cite{reger2019induction}. There, induction serves a goal that the user states; in GL, it is one of the methods tried on every conjecture.

\subsection{Admission and Rejection}
\label{sec:admission}

Existence statements create new objects. When an LB disintegrates $\exists z\colon P(\vec{x}, z)$, it can create a witness $w$ with $P(\vec{x}, w)$ (\cref{sec:compilation}). If every such statement created a witness, the new objects would feed rules that produce further existence statements, and the number of objects would keep growing: for addition alone, every pair of known numbers gives a new sum, and every new sum gives new pairs. GL therefore creates a witness only when something needs it. Each LB keeps two records that work as a pair:
\begin{itemize}
    \item \textbf{demands}, what the LB needs, written as a template with a marker $\square$ at the place of the missing object, for example $x + y = \square$, ``the value of $x + y$'';
    \item \textbf{parked sources}, existence statements whose object nothing has needed so far, filed under the template they would fill.
\end{itemize}
A new source looks for a matching demand, and a new demand looks for matching parked sources. Whichever comes second triggers the admission:
\begin{equation}
    \underbrace{x + y = \square}_{\text{demand}}
    \quad + \quad
    \underbrace{\exists z\colon x + y = z}_{\text{source}}
    \quad \Longrightarrow \quad
    x + y = w \qquad (w \text{ new}).
\end{equation}
A source without a matching demand is rejected: it is parked, not discarded, and its witness is created as soon as a matching demand appears. An admitted witness is never withdrawn. The two records thus form a duality: demands wait for sources, and rejected sources wait for demands.

\paragraph{Algebra.} On the algebra side, the hash engine applies rules. A rule whose premise needs a value that is not known yet files a demand for that value, and so does a goal. Sources are existence statements, such as the totality of addition, $\forall x, y\; \exists z\colon x + y = z$, instantiated at known numbers; for such a totality, GL builds the instance only when a demand for the value appears. For a rule $(a + b = c) \implies (b + a = c)$ and known numbers $a$ and $b$:
\begin{equation}
    \underbrace{a + b = \square}_{\text{demand of the rule}},
    \quad
    \underbrace{\exists z\colon a + b = z}_{\text{source}}
    \quad \Longrightarrow \quad
    a + b = w
    \quad \Longrightarrow \quad
    b + a = w.
\end{equation}
A value that no rule and no goal asks for is not created.

\paragraph{Integration.} On the integration side, GL assembles compound statements from their parts (\cref{sec:processor}). To prove a compound goal, such as an existence statement or a conjunction, it files a demand for every part that is still missing, and an object that fills such a part is admitted through the demand:
\begin{equation}
    \text{goal } \exists y\colon Q(\vec{x}, y) \wedge R(y)
    \quad \Longrightarrow \quad
    \text{demand } Q(\vec{x}, \square)
    \quad \Longrightarrow \quad
    \text{witness } w \text{ with } Q(\vec{x}, w).
\end{equation}
The construction then continues with the other parts, here $R(w)$. The strong induction step of \cref{sec:induction} uses the same mechanism: its demand $\square_1 + \square_2 = n$ admits a value below $n$, such as the remainder of a division. Sources that no integration needs are parked as on the algebra side and are revived when a matching demand arrives.

\paragraph{All mechanisms.} A new object, or a part of a construction, is admitted through one of these routes:
\begin{itemize}
    \item \emph{algebra demand}: a rule or a goal asks for a value, $x + y = \square$;
    \item \emph{operator input}: without a demand, the single input of an operator whose parameters are objects of the conjecture itself, not witnesses; for example an index $j$ with $F(j) = p$, that is, $F(\square) = p$;
    \item \emph{integration demand}: a compound goal asks for a missing part, $Q(\vec{x}, \square)$;
    \item \emph{statement fill}: a known statement $Q(\vec{x}, c)$ fills the demand $Q(\vec{x}, \square)$, and no new object is created;
    \item \emph{double demand}: two places are open at once, $\square_1 + \square_2 = n$, as in the strong induction step;
    \item \emph{negated universal}: a negated universal statement $\neg\forall x\colon \bigl(P(x) \implies Q(x)\bigr)$ gives a witness $x$ with $P(x) \wedge \neg Q(x)$, without a demand but within the depth limit below;
\end{itemize}
A case split admits no object, but it is opened on demand in the same way: an OR $E_1 \lor E_2$ is split into cases only when it is a premise of the task or a rule needs one of its disjuncts (\cref{sec:branching}).
Work that cannot go on yet waits instead of being discarded:
\begin{itemize}
    \item \emph{parked sources} wait for a demand. A new demand revives them, and so does an equality class that makes a parked template equal to a demand: a source parked under $x' + y = \square$ meets the demand $x + y = \square$ once $x' = x$ is known;
    \item \emph{parked subproofs}: the subproof of $A \implies B$ in a conjunction $G \wedge (A \implies B)$ opens only when $G$ is known;
    \item \emph{trial scopes}: when every demand that admits a new witness still needs further parts of its construction, the facts of the witness stay in a trial scope until those parts are known; only then is the witness admitted in the parent scope;
    \item \emph{limits}: a witness made from premises of generation at most $g$ has generation $g + 1$, and the configuration caps the generation, $\mathrm{gen}(w) \leq g_{\max}$; witnesses of integrations have a depth limit of the same kind;
    \item \emph{induction gates}: in an induction step, a demand must involve $\mathit{rec}$, $n$, the predecessor of $\mathit{rec}$, the successor of $n$ or an object admitted through such a demand (\cref{sec:induction});
    \item \emph{goal gates}: outside induction steps, an integration demand is filed only when it involves a variable of its goal or an object admitted through such a demand.
\end{itemize}

Other provers limit the growth of their search in related ways. Simplify and Z3 instantiate a quantified formula only with substitutions under which one of its patterns matches a known ground term, up to the known equalities; this is called E-matching \cite{detlefs2005simplify,demoura2007efficient}. Z3 instantiates the quantifiers that have helped to close branches at once and delays the others to a later round, which it uses when other options are exhausted \cite{demoura2007efficient}; parked sources in GL wait in a similar way, until a demand matches them. The set-of-support strategy allows no inference between two clauses outside a chosen set, usually the negated goal, so that every inference is connected to that set \cite{wos1965efficiency}. Magic sets rewrite the rules of a logic program so that bottom-up evaluation derives fewer facts that are irrelevant to the query \cite{bancilhon1986magic}. In GL, demands play a similar role, but only for new objects; rules fire on known statements without such a restriction.

\section{Hardware: the MPU}
\label{sec:mpu}

\begin{figure}[htbp!]
    \centering
    \includegraphics[width=\textwidth]{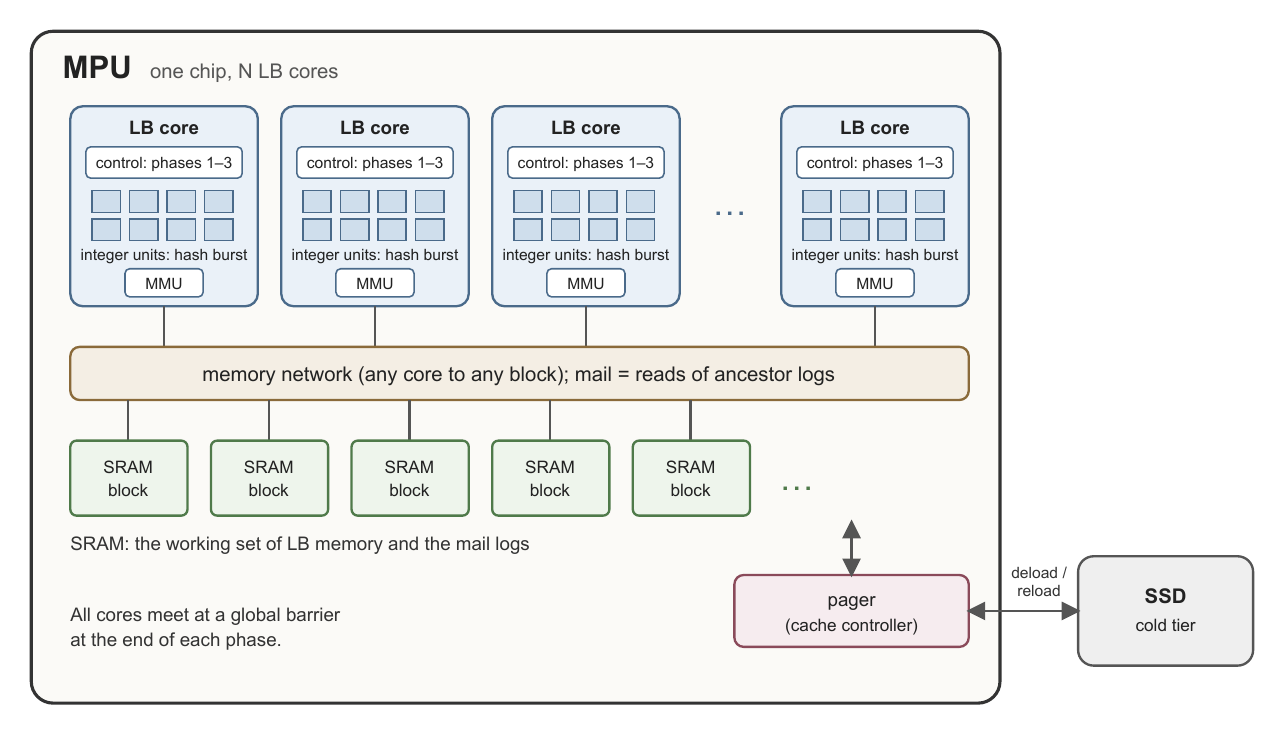}
    \caption{Overview of the MPU. LB cores with arrays of integer units and an MMU each reach the SRAM blocks through a memory network; mail is a read of the ancestors' logs. A pager keeps the working set of LB memory in SRAM and moves the rest to an SSD.}
    \label{fig:mpu}
\end{figure}

GL is designed together with a hardware architecture, the Mathematical Processing Unit (MPU), which is to be built as an application-specific integrated circuit (ASIC). The software already follows the constraints of such a chip: the prover keeps the data of its proof search in fixed, statically allocated pools, the three phases of its LB step allocate no dynamic memory, and every step is deterministic. One structure still uses dynamic memory: the registry of compiled definitions, which the phases of the LB step only read. Apart from this registry, the LB step of the software is therefore a working model of the hardware. This section outlines the architecture (\cref{fig:mpu}). A first test on a field-programmable gate array (FPGA) is part of the outlook (\cref{sec:outlook}).

Hardware for logical inference has been built before. The Parallel Inference Machine of the Japanese Fifth Generation Computer Systems project was designed to connect more than 100 processing elements, in clusters of eight that share memory, and to run the parallel logic language KL1 \cite{goto1988overview}. The Warren Abstract Machine defines an instruction set for Prolog that was also meant for dedicated processors \cite{warren1983abstract}. Both are designed for logic programs, in which the program determines the search. The MPU executes GL's hash bursts on definitions and proved theorems, and its unit of parallel work is the LB.

A proof engine is limited by memory and by time: the state of a problem may not fit into memory, or the deductions may take too long. The MPU addresses the first limit with a memory hierarchy that extends to a solid-state drive (SSD), and the second with parallel execution units.

\paragraph{LBs on cores.} A core executes one LB at a time, with its own control for the three phases of its step. The LBs far outnumber the cores: they pass through the cores in turn, and the pager keeps only the working set in SRAM (below). Its inference engine is an integer datapath: it builds hash keys, compares them and emits records; it needs no floating-point unit. Known statements are fixed-size integer records.

\paragraph{Memory.} The memory of the LBs is a hierarchy of one pool, blocks, pages and an arena per LB. In hardware, a block is a static random-access memory (SRAM) macro with its own port, so that many cores can access memory at the same time. An LB addresses its data by virtual offsets, which a small memory management unit (MMU) per LB translates into physical addresses. A pager acts as a cache controller: it keeps the working set of LBs in SRAM and moves the other LBs to an SSD and back. This deload and reload decouples the size of the SRAM from the size of the problem.

\paragraph{Mail.} Each LB writes its results once into its own log. In the first phase of its step, an LB reads the new entries of the logs of its ancestors. The tree of LBs is thus a pattern of memory reads over a shared memory, not a wiring between cores.

\paragraph{Parallel hash burst.} The hash burst of an LB splits into independent parts, which many small execution units can run at once. The results are merged by a sort, so the outcome does not depend on the order of execution. The prover has a second implementation of the hash burst that runs these parts on a graphics processor. Over a full run, the total time of the hash bursts is 27 times shorter on it than on the 32 threads of the processor route: 88 against 2,391 seconds, measured under Linux on the laptop of \cref{sec:case_study}. Almost all of the gain comes from the FTA batch, in which few LBs are active at a time; in the Incubator batches, with many active LBs, the two routes take about the same time. The two routes prove the same theorems; their proofs may differ. In the MPU, a core is therefore not one sequential unit but an array of small integer units, and the number of these units is the main lever for the run time. An LB with much work is split across several such arrays. Rule evaluation on graphics processors is also used for Datalog, over hash-indexed relations \cite{sun2025optimizing}.

\paragraph{Synchronization.} All cores meet at a global barrier at the end of each phase. Because every step is deterministic, the results do not depend on the timing of the cores.

\section{Verifiable Output}
\label{sec:output}

The results of GL are meant to be checked without trust in its prover. A run therefore ends with three outputs: a book of linked HTML pages for human readers, the report of an independent verifier, and an export to Lean~4 for a check by the Lean kernel. All three read the processed proof graph: the proof history of the run, reduced to the steps that each proof needs and ordered for reading. What a reader still has to trust is stated in \cref{sec:lean}.

The processed proof graph has the same basic form as a derivation in the TPTP language, the format of the TPTP library of problems for automated provers \cite{sutcliffe2017tptp}: a directed acyclic graph in which every derived formula names its inference rule and its parent formulas \cite{sutcliffe2006using}. In GL, the inference rule of a step is its proof tag. A system whose proofs can be checked independently by a small program satisfies the de Bruijn criterion \cite{barendregt2005challenge}. For the main batches, the Lean kernel is such a program. The verifier is an independent second check: a larger program that shares no code with the prover.

\subsection{The Proof Index}
\begin{figure}[htbp!]
    \centering
    \includegraphics[width=\textwidth]{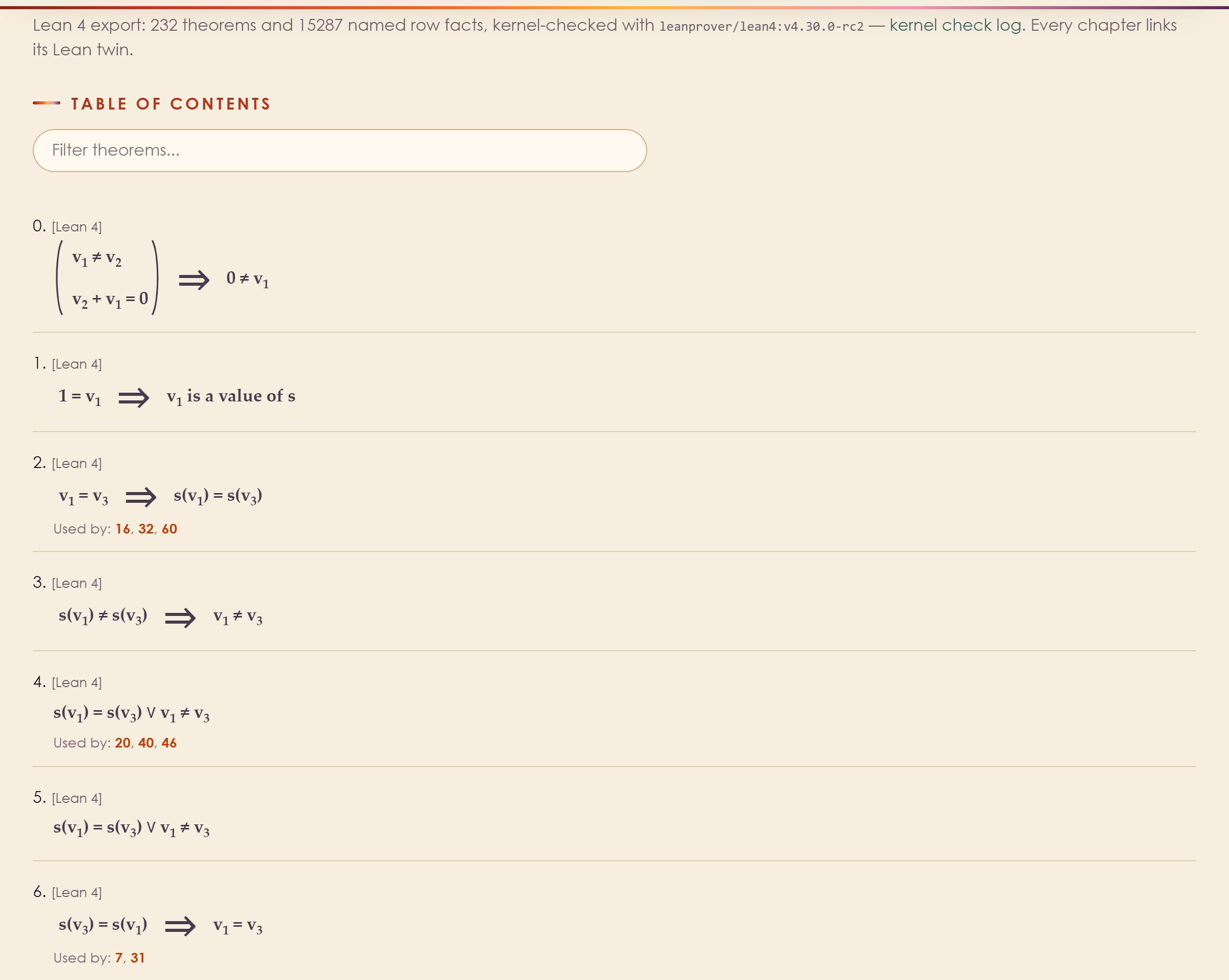}
    \caption{The proof index in common notation. Each entry links to its chapter and to its Lean version; ``Used by'' lists the chapters that cite the theorem.}
    \label{fig:index_screenshot}
\end{figure}

The entry point is the index page (\cref{fig:index_screenshot}). It lists every proved theorem in common mathematical notation, with links to the chapter that proves it, to its Lean version and to the chapters that use it. A switch shows the formal MPL expressions next to the common notation.

\subsection{The Proof Chapters}
\begin{figure}[htbp!]
    \centering
    \includegraphics[width=\textwidth]{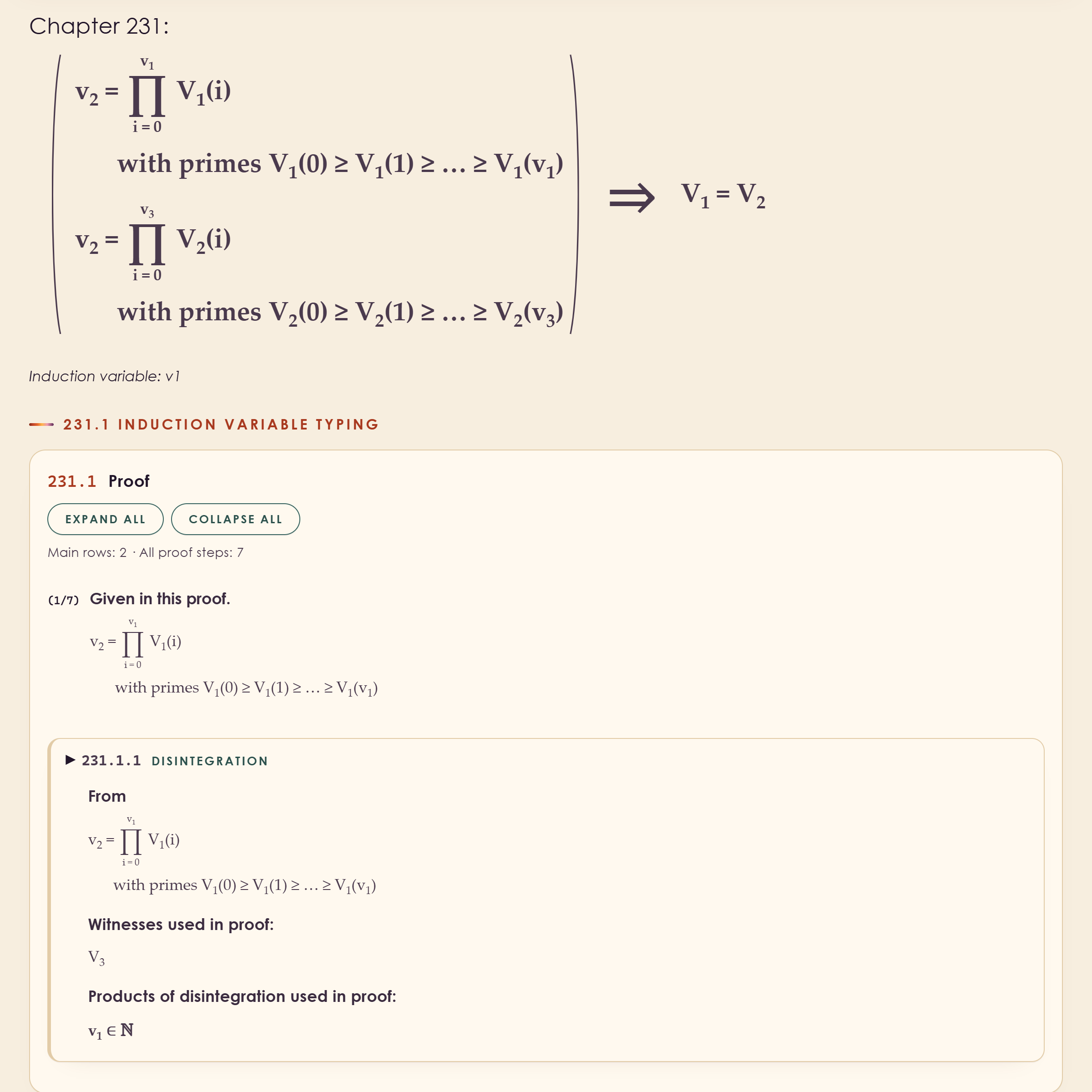}
    \caption{The proof page of the uniqueness part of the FTA in common notation: two nonincreasing sequences of primes with the same product are equal.}
    \label{fig:proof_chapter_screenshot}
\end{figure}

Each theorem has a chapter, one page with its complete proof (\cref{fig:proof_chapter_screenshot}). Every step shows the derived statement and its justification, such as an implication, a disintegration or an induction step, and links to the statements and theorems it uses. A reader can thus follow every proof back to the definitions. Steps that belong together, such as the unfolding of a definition, form blocks that open and close, and the page counts its main rows and all its proof steps. A reader does not need to trust GL: each step can be checked against its justification.

\subsection{The Verifier}
The verifier is an independent program. It is written in Python and shares no code with the prover; it shares a few routines with the Lean export, among them the parser of MPL expressions. It reads the processed proof graph, the GL binaries, that is, the compiler's table of the named parts of definitions, and one configuration file with the types and the input and output places of all operators; it does not read the MPL files. Like the Lean export, it therefore relies on the compiler for the meaning of the definitions (\cref{sec:lean}). It checks every proof step with a validator for its proof tag; there are 37 tags, among them implication, expansion, disintegration, equality, induction, contradiction and the tags of case splits. Further checks cover the whole graph: every cited source exists, every variable is bound once, the dependencies between theorems contain no cycle, and every theorem has its chapter. The verifier reports the number of checks and failures per tag, and any failure makes it end with an error. \Cref{sec:case_study} gives the counts for the case study.

\subsection{Export to Lean 4}
\label{sec:lean}
The proofs of the main batches are also translated into Lean~4 \cite{demoura2021lean4}, and the Lean kernel checks them. The export is a stage of the run: it starts when the processed proof graph is complete, the kernel check runs while the HTML pages are written, and a failed check fails the run.

\paragraph{The GL binary.} The export reads the processed proof graph, the configuration files and the GL binary of each batch; it does not read the MPL files. The GL binary is the main output of the compilation (\cref{sec:processor}). It lists every named part of the definitions, such as an existence statement, a conjunction or an implication, with its category, its arguments and its elements, together with the rules that the compiler derives from it. The proof steps cite these named parts, so the binary is what makes a translation step by step possible: every entry that the proofs use becomes one Lean definition with the same name and the same arguments. For the uniqueness part of the FTA, the binary contains the entry (shortened)
\begin{lstlisting}[language=MPL, numbers=none]
"primeFactorSeq": {"arity": 10, "category": "and",
  "elements": ["(sequence[u_1,u_3,u_5,u_8,u_9])",
               "(primeseq[u_1,u_3,u_4,u_6,u_7,u_9])",
               "(ordered[u_1,u_3,u_9])",
               "(fold[u_1,u_2,u_3,u_4,u_9,u_5,u_8,u_10])"], ...}
\end{lstlisting}
which becomes the Lean definition (binders shortened)
\begin{lstlisting}[language=MPL, numbers=none, mathescape]
def gl_primeFactorSeq (u_1 : GLSet $\alpha$) ... (u_10 : $\alpha$) : Prop :=
  gl_sequence u_1 u_3 u_5 u_8 u_9 $\wedge$ gl_primeseq u_1 u_3 u_4 u_6 u_7 u_9 $\wedge$
  gl_ordered u_1 u_3 u_9 $\wedge$ gl_fold u_1 u_2 u_3 u_4 u_9 u_5 u_8 u_10
\end{lstlisting}
Here $u_9$ is a sequence with the last index $u_8$, $u_{10}$ is its product, and the other arguments are the structure of the anchor. The binary is a text file in the JSON format. It is harder to read than MPL, but a person can read it, and the generated Lean definitions can be read as well.

\paragraph{Types.} The exported theory has one carrier type $\alpha$. A set is a predicate $\alpha \to \mathrm{Prop}$, and binary and ternary relations are predicates with two and three arguments. Every definition and every theorem is generic in the carrier; no GL object becomes a Lean type.

\paragraph{Certificate.} The translation has two stages. The first builds a certificate that does not depend on Lean: the typed expressions, the proof steps in an order in which every step comes after the steps it cites, their dependencies, and the SHA-256 hashes of the source files. It checks that every cited step and theorem exists and that the dependencies between theorems contain no cycle. The second stage writes the certificate as a Lean project.

\paragraph{One Lean fact per step.} Each proof step becomes one named Lean fact, whose comment names the GL rule and the source step, and its proof does what the proof tag of the step says. An implication step applies the cited rule to the cited premises in the order of the rule. A chain of two equalities is composed with \texttt{Eq.trans}, and a rewriting by an equivalence class eliminates the cited equalities with Lean's recursor for equality. A contradiction step applies the cited negation to the cited statement. A case split unfolds the compiled OR and applies the result of each case. The witness of an existence statement is obtained once and shared by all steps about it. The export searches for no proof. The proofs consist of explicit terms and of basic tactics that introduce assumptions, apply the cited facts and split cases; definitions are unfolded with \texttt{simp only} and a list of named definitions and laws. No tactic looks for a lemma, and every step uses the facts that its GL step cites. The writer rejects \texttt{sorry}, \texttt{admit}, \texttt{axiom}, \texttt{opaque} and \texttt{unsafe}.

\paragraph{Induction and the natural numbers.} The exported theorems take the anchor and two induction principles, for ordinary and for strong induction, as premises, so they hold for every structure that satisfies these premises. The induction parts of GL (\cref{sec:induction}) apply these principles explicitly. A model file proves the anchors for Lean's natural numbers, obtains the two principles from \texttt{Nat.rec} and \texttt{Nat.strongRecOn}, and applies every theorem to them; it uses Lean's tactics \texttt{simp} and \texttt{omega} for the anchors. A test confirms that the induction premise is not trivial: a structure with the double successor in place of the successor does not satisfy it. On the natural numbers, the two parts of the FTA read, with the arguments of the anchor abbreviated to $\ldots$:
\begin{lstlisting}[language=MPL, numbers=none, mathescape]
theorem fta_source_228 : $\forall$ (v1 : Nat), gl_preorder ... 2 v1 $\to$
    gl_hasPrimeFactorization ... v1
theorem fta_source_231 : $\forall$ (v1 : Nat) (V1 : GLBinaryRelation Nat) (v2 : Nat),
    gl_primeFactorSeq ... v1 V1 v2 $\to$
    $\forall$ (v3 : Nat) (V2 : GLBinaryRelation Nat), gl_primeFactorSeq ... v3 V2 v2 $\to$ V1 = V2
\end{lstlisting}
The first states that every $n \geq 2$ has a factorization into a nonincreasing sequence of primes; the second, that two such factorizations of the same number are equal.

\paragraph{What the reader trusts.} The Lean kernel checks every step of the export. For all 232 exported theorems on the natural numbers, \texttt{\#print axioms} reports only standard axioms of Lean: \texttt{propext}, \texttt{Quot.sound} and, for 189 of them, \texttt{Classical.choice}; the export adds none. The kernel does not check that the exported definitions mean what the MPL definitions mean. The translation from MPL to the binary is done by GL's compiler, and the reader relies on it, as on the definitions themselves. Since the binary and the generated definitions can be read, a reader can also check this translation, definition by definition. The statements use GL's definitions, such as \texttt{gl\_primeFactorSeq}, not those of Mathlib, the mathematical library of Lean \cite{mathlib2020lean}.

\begin{figure}[htbp!]
    \centering
    \includegraphics[width=\textwidth]{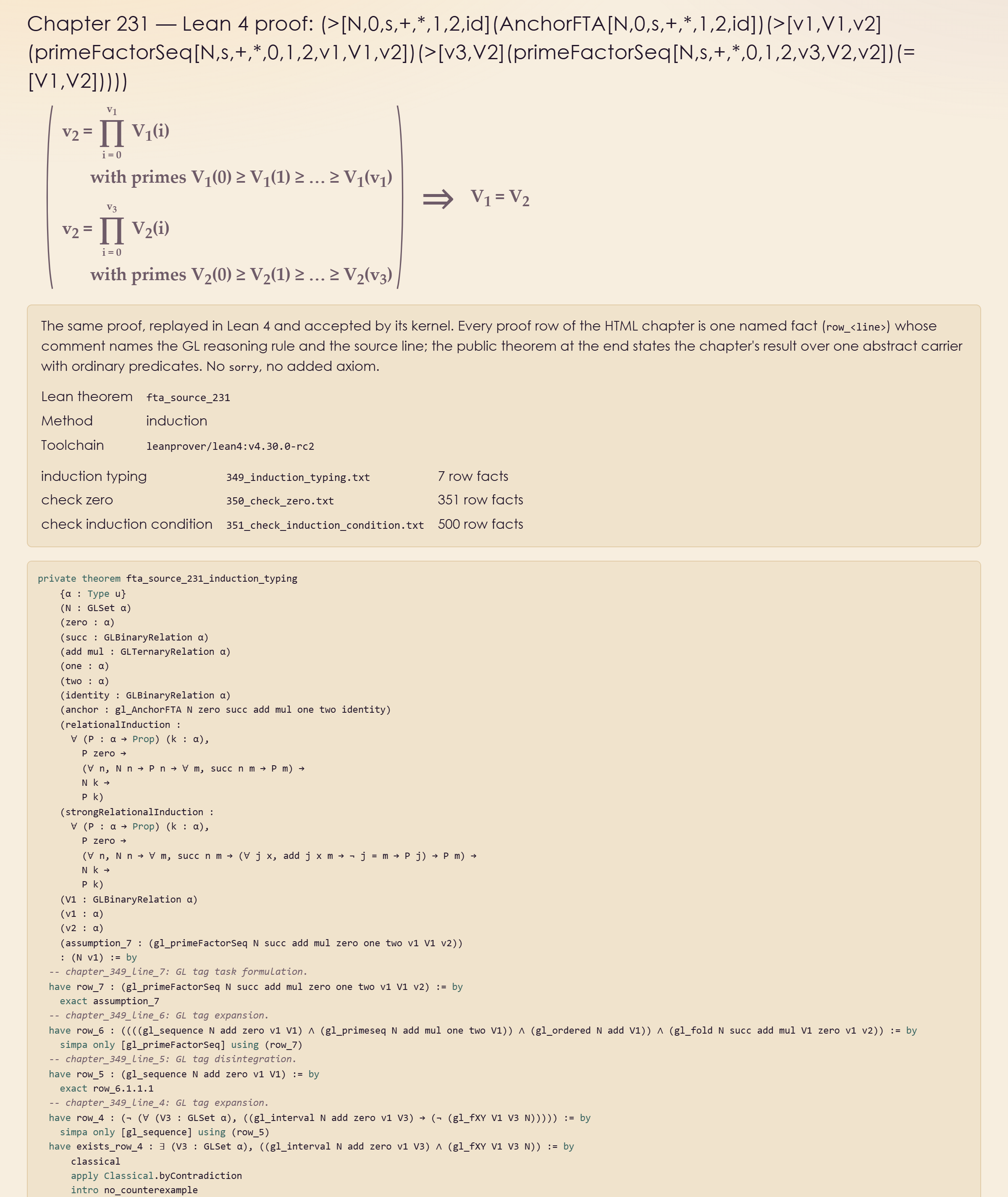}
    \caption{The Lean~4 page of the same proof. Each proof step of the HTML page is one named fact in Lean, and the Lean kernel checks the whole proof.}
    \label{fig:lean_chapter}
\end{figure}

\paragraph{Size.} For the run of \cref{sec:case_study}, the export has 183 definitions and 232 theorems with 15,287 proof steps, and the kernel check takes 37 seconds. Each HTML chapter links to its Lean version (\cref{fig:lean_chapter}).

\section{Case Study: Elementary Number Theory}
\label{sec:case_study}

The case study develops elementary number theory from the definition of the natural numbers. A full run consists of eight batches: two Incubator batches and the main batch for Peano arithmetic, three Incubator batches and the main batch for Gauss's summation formula, and one batch with the supplied conjectures for the FTA. On a laptop workstation (Dell G16 7630) with an Intel Core i9-13900HX processor (24 cores, 32 threads), 32~GB of RAM and an NVIDIA GeForce RTX~4070 Laptop GPU, which runs the hash bursts, under Windows~11, a full run takes 5.8 minutes. The run uses release v0.13.0 (\cref{sec:availability}) with its configuration files, built with Visual Studio 2022 (MSVC 14.44) and CUDA 13.3; the Lean kernel check uses Lean 4.30.0-rc2. \Cref{tab:runtime} gives the run time of each batch.

\subsection*{Peano Arithmetic}

The conjecturer produces 1,001 candidates. The batch also receives one supplied lemma, $c + s(a) = s(b) \implies c + a = b$, which the proofs of the FTA use; no proof of a Peano or Gauss theorem in the proof graph cites it. The CE filter passes 172 of the 1,002 conjectures to the prover. The prover runs 22 iterations and proves 129 theorems, which the compressor reduces to 36 essential ones. Of the 172 conjectures that passed the filter, GL proves 120 as stated and 7 in a stronger form with fewer premises (\cref{sec:scopes}); 45 remain open. The proved theorems include the commutativity and associativity of addition and multiplication and the distributive law.

\subsection*{Gauss's Summation Formula}

The Gauss batch adds intervals, sequences and the fold operator to the definitions and generates 673 conjectures, and the CE filter passes 80 conjectures to the prover. Its configuration restricts the conjecturer strongly: it admits a fixed list of expressions, each at most a given number of times in a conjecture, for example at most one \texttt{fold}; it limits a conjecture to at most 4 simple expressions; and it excludes a list of patterns and of combinations of expressions. It does not contain the summation formula: the formula is one of the conjectures that the conjecturer generates within these limits. The anchor of the Gauss batch is connected to the anchor of the Peano batch, so the essential Peano theorems are available as proved knowledge; the prover makes them available to every LB before the proof search. The prover runs 35 iterations and proves 40 theorems, among them Gauss's summation formula. Of the 80 conjectures that passed the filter, GL proves 18 as stated and 1 in a stronger form; 61 remain open. GL derives the formula in a division-free form, with the product of $n$ and its successor equal to twice the sum:

\begin{equation}
    n(n+1) = 2\sum_{i=1}^{n} i
\end{equation}

The compressor reduces these theorems and the essential Peano theorems to 63 essential theorems.

\subsection*{The Fundamental Theorem of Arithmetic}

A further batch receives the definitions of the FTA and a supplied list of 103 conjectures. The list contains the two parts of the FTA themselves: every $n \geq 2$ is the product of a nonincreasing sequence of primes (existence), and two such sequences with the same product are equal (uniqueness). The FTA batch starts from the 63 essential theorems of the Peano and Gauss batches of the same run. GL proved all of them and conjectured all but the supplied lemma of the Peano batch; the proofs of the FTA cite 33 of them, the supplied lemma among them. GL proves all 103 supplied conjectures, both parts of the FTA included (\cref{fig:proof_chapter_screenshot}). It receives neither their proofs nor the order in which to prove them, nor hints for their use.

\paragraph{The list.} The list was written during the development of the proof: the author wrote a proof by hand for each stage, and AI agents wrote its lemmas as conjectures in MPL (\cref{sec:agents}). The final list was then derived from GL's own proof graph. It contains every theorem that the proofs of the two parts cite, and every theorem that those proofs cite, down to the definitions. It leaves out what GL derives itself from a conjecture it is given, such as restated theorems and OR theorems. Two conjectures are added that no proof cites. The first, $v \in \mathbb{N} \implies 1 \mid v$, has facts that make GL request the expansion of a statement that the base case of an induction starts from. The second states that a sequence extended by one value and cut back to its old length is the original sequence; it shows GL that two of its objects are one.

\paragraph{Content.} The conjectures cover the order of the natural numbers, divisibility, a form of B\'ezout's identity without negative numbers, Euclid's lemma, primes and the least divisor, and finite sequences, which GL represents as functions on an interval $[0, k]$: their extension by one element, their products and their order. Most of them are natural steps that a textbook proof leaves to the reader, for example
\begin{equation}
    2 \leq v \implies v \neq 1, \qquad
    v_1 \mid v_2 \;\wedge\; v_2 \mid v_3 \implies v_1 \mid v_3,
\end{equation}
or that a prime that divides the product of a sequence of primes is one of its members. Many state a single step of an argument, such as that a sequence on $[0, k + 1]$ is a sequence on $[0, k]$ extended by its last value. A few are steps of particular proofs, such as the rearrangements of B\'ezout's equation in its induction step, and the two conjectures that no proof cites are on the list only because the search needs them.

\paragraph{Comparison.} The FTA is number 80 on Wiedijk's list of 100 theorems and their formalizations, which names twelve systems with a formal proof of it, among them HOL Light, Isabelle, Lean, Rocq, Mizar and ACL2 \cite{wiedijk2026formalizing}. In the proof assistants among them, a person writes the proof and the system checks it. The Boyer--Moore prover proved the theorem in 1979 from a sequence that its authors supplied in the order of proof \cite{boyer1979computational,boyer1979unique}. The part for the FTA begins with the definition of a prime and has 9 definitions and 41 lemmas; it builds on the arithmetic and list theorems that the authors supplied earlier and that the prover proved from the Peano axioms. Every theorem that later proofs may use carries its declared use, for example as a rewrite rule. Their statement differs from the one here: a factorization is a list of primes, and uniqueness holds up to the order of the factors. The 1979 theory has no quantifiers: an existence statement is replaced by a function that computes the object, which the authors define, here a function that computes the factors of every nonzero number. Its definition principle admits only functions whose recursion terminates; the definition of a prime, for example, considers only the numbers below it as divisors. On top of the Peano axioms, MPL defines concepts without recursion, with quantifiers (\cref{sec:mpl}), and GL proves existence statements and builds them from their parts (\cref{sec:processor,sec:admission}). Most of their lemmas are small steps, as in GL's list, and some are named as helpers, such as \texttt{HACK1} and \texttt{KLUDGE.BRIDGE}. GL receives its 103 conjectures as a set, without an order and without declared uses; it proved all the Peano and Gauss theorems that its proofs cite, and it conjectured all of them but the one lemma supplied to the Peano batch. Its proofs are checked by the verifier and by the Lean kernel. We know of no earlier system that assembles a proof of the FTA from supplied lemmas without their order and without hints.

\subsection*{Incubator}

Before each main batch, Incubator batches generate its ground facts. The Incubator for Peano arithmetic uses an anchor that names nine numerals, $0, 1, \ldots, 8$; the natural numbers themselves are not restricted to these nine. It proves the right value of an operator as an implication and refutes the false candidates by contradiction (\cref{sec:ce_filter}). The Incubator batches process 2,035 conjectures and prove 1,462 of them. The operator facts obtained from the proved implications form successor, addition and multiplication tables, for example $s(0) = 1$, $0 + 5 = 5$ and $2 \cdot 3 = 6$; the proved inequalities include $s(0) \neq 4$ and $0 \neq 1$. The tables stop at the named numerals: $5 + 4$ has no named value, so the Incubator proves the nine inequalities $5 + 4 \neq c$ for $c = 0, \ldots, 8$, but no fact $5 + 4 = c$. More named numerals would extend the tables at the cost of a larger search. These facts feed the CE filter of the main batches.

The Incubator is heuristic in one respect. It reformulates a proved statement of the form $2 + 2 = x \implies x = 4$ into the operator fact $2 + 2 = 4$. Strictly, this step omits the proof that $2 + 2$ has a value at all, which matters for compound operators such as \texttt{fold}. The risk is therefore not that a false fact enters the proof graph, but that the CE filter rejects a true conjecture because of an erroneous Incubator fact, so the existence of such values has to be ensured in advance. The proof graph may thus miss some valid theorems. Incubator facts never enter a proof of the main batches, and every theorem in the proof graph is checked by the verifier, those of the main batches also by the Lean kernel.

\subsection*{Compressor}

The compressor (\cref{sec:compressor}) passes only the essential theorems of each batch to the later batches; every removed theorem remains derivable from them.

\subsection*{Verifier and Lean}

The verifier (\cref{sec:output}) performs 265,663 checks on all proof chapters, 198,961 for the Incubator and 66,702 for the main batches, and reports no failure. The Lean kernel accepts the export of all 232 theorems of the main batches.

\begin{table}[ht]
\centering
\caption{Run time in seconds by batch on the laptop workstation. The prover column includes the hash bursts, which run on the GPU. Total is the time of the conjecturer and of the batch's executable, which runs the CE filter, the prover and the compressor. The processing of the proofs, the HTML, the verifier and the Lean export run once per run and are included only in the full-run time. ---: the batch has no such stage.}
\label{tab:runtime}
\begin{tabular}{lrrrr}
\hline
Batch & Conjecturer & CE filter & Prover & Total \\
\hline
Incubator Peano 1 & 1.9  & ---  & 51.2 & 59.5 \\
Incubator Peano 2 & 0.5  & ---  & 7.0  & 8.6 \\
Peano             & 41.8 & 4.1  & 8.4  & 60.0 \\
Incubator Gauss 1 & 0.9  & ---  & 27.7 & 34.2 \\
Incubator Gauss 2 & 0.2  & ---  & 6.5  & 8.4 \\
Incubator Gauss 3 & 0.3  & ---  & 1.4  & 2.1 \\
Gauss             & 5.2  & 3.8  & 6.1  & 18.0 \\
FTA               & ---  & ---  & 78.6 & 83.3 \\
\hline
\multicolumn{5}{l}{Verifier: 265,663 checks, 0 failures; Lean kernel check passed} \\
\hline
\multicolumn{5}{r}{Full run, wall-clock: 348 seconds} \\
\hline
\end{tabular}
\end{table}

\subsection*{Observations}

Of the 348 seconds of a full run, the Incubator batches take 113 and the FTA batch 83. The cost of incubation is incremental: ground facts do not need to be regenerated on every run, a new definition layer incubates only its own facts on top of the existing ones, and previously incubated facts can be supplied externally.

The Incubator output also shows a further capability: GL not only proves properties but also computes concrete results from the definitions. The next section describes this capability.

\section{From Proofs to Calculations}
\label{sec:calculation}

The Incubator results show a further capability of the architecture: GL performs concrete numerical calculations. The conjecturer of the Incubator proposes every named numeral as a candidate value, and the prover proves the implication for the right one, for example that $2 + 3 = x$ implies $x = 5$, or that $\text{fold}(+, [0 \ldots 3]) = x$ implies $x = \sum_{i=0}^{3} i = 6$. The value thus comes from the conjecturer, but the derivation obtains it from the axiomatic definitions of the operators: from $2 + 3 = x$ it derives $s(4) = x$, and from there $x = 5$. Every intermediate step is recorded in the proof graph with its provenance.

Three levels of calculation appear in the Incubator output:

\begin{enumerate}
  \item \textbf{Evaluation of operators.} GL evaluates the successor, addition and multiplication on concrete arguments, for example $s(2) = 3$, $4 + 1 = 5$ and $2 \cdot 3 = 6$, by following the recursive definitions to their base cases. The proved theorem has the form $4 + 1 = x \implies x = 5$.

  \item \textbf{Relations defined through operators.} GL decides such relations on concrete values, for example $2 \le 5$, which by definition means that $2 + p = 5$ for some natural number $p$.

  \item \textbf{Execution of multi-stage algorithms.} The \texttt{fold} operator describes an iterated accumulation over an interval, the logical equivalent of a \texttt{for} loop. When GL evaluates a \texttt{fold} instance, it establishes the interval bounds, passes the accumulator through each element and arrives at the final value. The proof graph records every iteration.
\end{enumerate}

There is no separate calculation mode. The same MPL definitions, the same inference engine and the same LB grid that prove abstract properties such as the commutativity of addition also compute that $2 + 3 = 5$. The proved implications therefore carry the same provenance as every other proved theorem: a chain of justified steps back to the axioms.

The operator facts themselves, such as $2 + 3 = 5$, are not proved theorems. The Incubator obtains them from the proved implications by back-reformulation, which omits the proof that the value exists (\cref{sec:case_study}); they feed the CE filter and never enter a proof of the main batches. This shortcut differs from the intended calculation path. The Incubator runs before the main batches: the theorems it would need for a fully rigorous derivation are not yet conjectured, let alone proved. A future calculator, by contrast, runs after the theorem base has been established. It uses the verified output of GL and calculates with proved rules only, without the reformulation step, so the calculation path is as rigorous as the proof path.

This opens a path toward a computer algebra system (CAS) with full provenance: a system in which every computed result is a theorem, with each intermediate step traceable to the axioms. The current capability is modest: single-digit arithmetic on the named numerals $0, \ldots, 8$.

\subsection{Toward a Full-Provenance CAS}

Such a CAS will begin with integers. Once integer arithmetic is established with full provenance, it will extend to fixed-point and floating-point numbers. The long-term goal is a calculator in which not only the method of calculation is founded on axioms, but every digit of the output carries a provenance graph back to the definitions.

Earlier work combines computer algebra with theorem proving. Harrison and Th\'ery let the CAS Maple search for a result and the prover HOL check it \cite{harrison1998skeptic}. Kaliszyk and Wiedijk built a prototype CAS on top of the proof assistant HOL Light, which proves every simplification that the CAS performs \cite{kaliszyk2007certified}. In GL, the value is computed by the same engine that proves theorems, and its derivation is its proof.

\section{Current Stage: FTA Discovery}
\label{sec:discovery}

The proof of the FTA in \cref{sec:case_study} is assembled from a supplied list of conjectures. Current work extends the conjecture generation, filtering and proving to the FTA, so that GL finds the required lemmas itself. The supplied conjectures of the assembled proof mark the target: the discovery has to produce at least these.

The approach is the one that works for Peano arithmetic and Gauss's summation formula, at a larger scale:
\begin{enumerate}
    \item The conjecturer generates enough conjectures that the lemmas the FTA needs are among them.
    \item The CE filter removes almost all of them.
    \item GL proves the remaining conjectures in successive rounds.
    \item After each round, the compressor reduces the proved theorems to their logical core, which keeps the memory small for the next round.
\end{enumerate}
The volume of processed information is expected to exceed the RAM; the SSD deload (\cref{sec:mpu}) is meant to hold it.

About 85\% of the work of this campaign is expected to go into accelerating GL. The rest goes into adjusting the existing discovery machinery, developed on Peano arithmetic and Gauss's summation formula, to the definitions of the FTA.

\section{Related Work and Comparison}
\label{sec:related}

GL belongs to the work on automated mathematical discovery: systems that generate conjectures and, in some cases, prove them. This section compares GL first with these systems and then with provers, which prove a goal that a user states. \Cref{tab:comparison} summarizes the comparison.

\subsection{Discovery Systems}

\paragraph{Conjectures without proofs.} AM \cite{lenat1976am,lenat1977automated} started from about 100 concepts of finite set theory and 250 heuristic rules. It defined new concepts, among them the natural numbers and the primes, and conjectured regularities from computed examples, among them unique prime factorization and Goldbach's conjecture. It proved nothing; proof was outside its design \cite{lenat1977automated}. EURISKO \cite{lenat1983eurisko} added heuristics that create new heuristics. Graffiti \cite{fajtlowicz1988conjectures} generates inequalities between invariants of graphs that hold for all graphs in its database; mathematicians proved or refuted many of them. Graffiti.pc \cite{delavina2005graffitipc}, AutoGraphiX \cite{caporossi2000variable}, the program of Larson and Van Cleemput \cite{larson2016automated} and TxGraffiti \cite{davila2026automated} follow the same pattern.

\paragraph{Conjectures with proofs.} HR \cite{colton1999automatic,colton2002automated} forms concepts and conjectures from examples in number theory and finite algebras, searches for counterexamples with the model generator MACE and proves conjectures with the first-order prover OTTER. MATHsAiD \cite{mccasland2017mathsaid} is closest to GL in its aim: it starts from user-supplied axioms and definitions and reports a conjecture as a theorem only after its own natural-deduction prover, which includes induction rules, has proved it. Theory exploration produces equational lemmas about recursive functions. QuickSpec \cite{claessen2010quickspec,smallbone2017quick} finds candidate equations by random testing, HipSpec \cite{claessen2013automating} proves them by structural induction with an automated first-order prover, and Hipster \cite{johansson2014hipster} proves them with Isabelle tactics, so that Isabelle's kernel checks the proofs. IsaCoSy \cite{johansson2011conjecture} and IsaScheme \cite{montanorivas2012scheme} synthesize conjectures in Isabelle, filter them with Isabelle's counterexample finders and prove them with Isabelle-based provers. TheSy \cite{singher2021theory} replaces random testing with a symbolic comparison of terms and proves the lemmas with its own prover. Theorema \cite{buchberger2000theory,buchberger2006theorema} supports the exploration of theories with its own provers; there, a human supplies most conjectures, and failed proofs suggest further lemmas. Theory exploration has the same shape as the GL pipeline: generate conjectures, discard those that counterexamples refute, and prove the rest by induction.

\paragraph{Data and machine learning.} Several recent systems find conjectures or constructions with numerical or learned methods. The Ramanujan Machine \cite{raayoni2021generating} matches continued-fraction formulas to numerical constants; its formulas come without proofs. Davies et al.\ \cite{davies2021advancing} guide the intuition of mathematicians with trained models; in knot theory, this led to a theorem that the mathematicians proved, and in representation theory to a conjecture. FunSearch \cite{romeraparedes2024mathematical} and AlphaEvolve \cite{novikov2025alphaevolve} evolve programs with language models and score them by running an evaluator, and Wagner \cite{wagner2021constructions} finds counterexamples with reinforcement learning; these systems themselves produce no formal proofs. Georgiev et al.\ combined AlphaEvolve with two further systems in one case: for the finite field Kakeya problem, AlphaEvolve found a general construction, Deep Think proved its correctness in natural language, and AlphaProof formalized the proof for three dimensions in Lean; for four dimensions, the proofs were too difficult for AlphaProof, and the authors checked them by hand. The authors passed the results from one system to the next \cite{georgiev2025mathematical}. In GL, the conjectures and their proofs come from one engine. AlphaGeometry \cite{trinh2024solving} generates synthetic geometry theorems whose proofs come from a symbolic engine, without a formal kernel. Minimo \cite{poesia2024learning} starts from axioms, among them those of Peano arithmetic: a language model proposes conjectures, and a learned prover proves them in its own environment. LeanConjecturer \cite{onda2025leanconjecturer}, STP \cite{dong2025stp} and Lemmanaid \cite{alhessi2025lemmanaid} generate conjectures for Lean or Isabelle with language models, mainly to train or support provers.

\subsection{Provers}

Provers start from a goal that the user states. Saturation provers such as E \cite{schulz2019e2.3} and Vampire \cite{kovacs2013vampire,bartek2025vampire} and SMT solvers such as Z3 \cite{demoura2008z3} and cvc5 \cite{barbosa2022cvc5} prove or decide single problems automatically. Interactive proof assistants such as Lean~4 \cite{demoura2021lean4}, with its library Mathlib \cite{mathlib2020lean}, and Isabelle/HOL \cite{nipkow2002isabelle} check proofs that humans write; Lean does so in a small trusted kernel. Hammers such as Sledgehammer \cite{paulson2012sledgehammer,blanchette2013smt} and CoqHammer \cite{czajka2018coqhammer} send a single goal to automated provers and reconstruct the proof in the assistant. The Boyer--Moore prover and its successor ACL2 \cite{boyer1979computational,kaufmann2000acl2} prove a sequence of user-supplied conjectures by induction, guided by hints for the use of each proved theorem; in this way the Boyer--Moore prover proved the unique prime factorization theorem in 1979 \cite{boyer1979unique}. Recent provers based on language models, among them AlphaProof \cite{hubert2026alphaproof}, DeepSeek-Prover-V2 \cite{ren2025deepseekproverv2}, Goedel-Prover \cite{lin2025godelprover,lin2026goedelproverv2}, Kimina-Prover \cite{wang2025kimina}, Seed-Prover \cite{chen2025seedprover} and Aristotle \cite{achim2025aristotle}, prove formal Lean statements, and Lean checks their proofs. Seed-Prover and Aristotle also propose lemmas, but only for the one goal they are given.

\subsection{Comparison}

GL belongs to the discovery systems that also prove. Within this group, it differs in four respects. It uses one deterministic engine for counterexample filtering, proof and calculation. It records every step with its justification and has every proof checked by an independent verifier and, for the main batches, by the Lean kernel; among the discovery systems of \cref{tab:comparison}, Hipster also has its proofs checked by a kernel, that of Isabelle. It works in theory layers, with compression to an essential core between them. And it is designed together with a hardware architecture. Like the Boyer--Moore prover, GL can build a proof of the FTA from supplied lemmas (\cref{sec:case_study}); unlike it, GL needs no hints for individual theorems, and the current work aims to let GL find the lemmas itself (\cref{sec:discovery}).

\begin{table}[htbp]
\centering
\small
\caption{Comparison of GL with discovery systems and provers, as described in the cited sources. ``---'': not applicable.}
\label{tab:comparison}
\begin{tabularx}{\textwidth}{>{\raggedright\arraybackslash}p{2.5cm} >{\raggedright\arraybackslash}X >{\raggedright\arraybackslash}X >{\raggedright\arraybackslash}X >{\raggedright\arraybackslash}X}
\toprule
System & Input & Conjectures and filter & Proof & Check of the proof \\
\midrule
AM & concepts and heuristics & from examples; interestingness & none & --- \\
Graffiti and successors & objects and invariants & inequalities that hold on all stored objects & by humans & --- \\
HR & concepts and examples, or axioms & from examples; counterexample search & OTTER & none reported \\
MATHsAiD & axioms and definitions & generated; triviality filters & own prover with induction & none reported \\
HipSpec, Hipster & functional programs or Isabelle theories & equations; random testing & induction & Hipster: Isabelle's kernel \\
TheSy & datatypes and recursive functions & equations; symbolic comparison & own prover with induction & none reported \\
Minimo & axioms & from a language model; difficulty & tree search & no separate check \\
FunSearch, AlphaEvolve & an evaluation function & programs and constructions; evaluator & none & --- \\
Boyer--Moore, ACL2 & definitions, conjectures and hints & none & induction prover & none reported \\
LLM provers & a formal goal & lemmas for that goal only & language model & Lean \\
GL & definitions and configuration & generated, or supplied; counterexamples from Incubator facts & one deterministic engine & verifier and Lean kernel \\
\bottomrule
\end{tabularx}
\end{table}

\subsection{Contributions}

\begin{itemize}
    \item \textbf{One engine for discovery and proof.} GL generates conjectures from definitions, removes those that counterexamples from the ground facts of the Incubator refute, and tries to prove the rest with one deterministic, hash-based engine that also calculates.
    \item \textbf{Checked output.} Every proof step carries its justification; an independent verifier and the Lean kernel check the proofs.
    \item \textbf{Reasoning mechanisms in a parallel grid.} Scopes, equivalence classes, case splits, contradiction and induction work inside a grid of independent LBs.
    \item \textbf{Hardware co-design.} The software follows the constraints of a chip and runs its hash bursts on a GPU; the MPU maps it to silicon.
\end{itemize}

\section{Integration with an AI Agent}
\label{sec:agents}

Whether an AI agent can operate GL comes down to whether it can write MPL files and configuration files. By now, AI agents have written all of GL's MPL and configuration files, and they maintain them. They are also deeply involved in coding, debugging and problem analysis. In 2025, it was an open question whether AI agents could operate GL; today the answer is a plain yes. One reason is that MPL is, at its core, a small subset of common mathematical notation, which AI agents handle easily.

In the other direction, an AI agent can use GL as a tool. When the agent faces a mathematical reasoning task, and in the future any task that can be expressed in MPL, it can translate the task into MPL, let GL derive the consequences deterministically, and use the checked results in its further work. GL stays an external proof engine: the agent runs it as a separate program, not as a library inside its own code.

Related work connects language models with proof assistants. Translating mathematics automatically into the language of a proof assistant is called autoformalization; Wu et al.\ study it with language models for Isabelle/HOL \cite{wu2022autoformalization}. LeanDojo lets programs interact with the proof assistant Lean; its prover, ReProver, is a language model that selects premises from Lean's mathematical library by retrieval \cite{yang2023leandojo}. There, the language model proposes the proof steps, and Lean checks them. With GL, the agent writes the task in MPL, and GL derives the consequences and their proofs.

\section{Outlook}
\label{sec:outlook}

The next steps are:
\begin{enumerate}
    \item \textbf{FTA discovery.} GL is to find the lemmas of the FTA itself, without supplied conjectures (\cref{sec:discovery}).
    \item \textbf{FPGA.} The prover code is to be ported to an FPGA, and a single LB is to be executed on it as an experiment. This is the first test of the MPU design (\cref{sec:mpu}) in hardware.
    \item \textbf{Hardware verification.} The first practical application is to be hardware verification: proofs that a hardware design, for example an arithmetic circuit, meets its specification.
    \item \textbf{Algebraic number theory.} The next mathematical target is algebraic number theory. It needs a feature that GL does not have yet: sets as elements of other sets. For example, ideals are sets of numbers, and the theory works with sets of ideals. The type notation of MPL can already write such a type, but the conjecturer, the compiler and the Lean export support only four types today: elements, sets, and binary and ternary relations. All three have to be extended, not only the prover. The theories so far did not need this feature, so it has not been developed. A first step in this direction exists: GL proves an equality of sets differently from an equality of numbers. It proves $X = Y$ by showing that $x \in X \Leftrightarrow x \in Y$ for every $x$.
    \item \textbf{Cryptography.} After algebraic number theory, cryptography is planned as the next application; much of it rests on number theory.
\end{enumerate}

GL takes over the Peano axioms, whose equations for addition and multiplication are recursive. Every definition on top of them is non-recursive: no MPL definition refers to itself, and concepts such as divisibility, primes, finite sequences and their products are defined by axioms with quantifiers over elements, sets and relations. MPL therefore admits definitions that no terminating recursion computes, such as the real numbers as a complete ordered field, whose completeness axiom quantifies over all subsets as the induction axiom of the second-order Peano axioms does, and with them notions of analysis such as continuity. Whether the conjecture generation and the proof search of GL carry over to such theories has not been tested, and topology in general needs the sets of sets named in step 4.

\section{Conclusion}
\label{sec:conclusion}

GL set out to answer a narrow question: whether a deterministic architecture that starts from definitions can generate and prove theorems without human guidance at the level of individual proof steps.

The answer is yes for the first two parts of the case study (\cref{sec:case_study}): from supplied definitions, GL derives the basic laws of Peano arithmetic and Gauss's summation formula. For the FTA, GL so far builds the proof from previously proved lemmas, further definitions and a supplied list of conjectures; it receives the conjectures, but not their proofs or the order in which to prove them. The independent verifier checks every proof step and reports no failure, and the Lean kernel accepts the exported proofs of the main batches.

Three properties of the architecture held in practice. First, rules fire by hash-table lookups, and every proof step, also those of equivalence classes, induction and case splits, is recorded with its justification, so every derivation can be replayed and checked: the proof graph is a complete record, not a summary. Second, each LB holds only its own part of the task and receives the results of other LBs only through mail, in the first phase of its step. The LBs can therefore work in parallel; the hash bursts already run on a graphics processor, and the MPU design maps the same structure to hardware. Third, in the batches for Peano arithmetic and Gauss's summation formula, GL was given definitions and one lemma, which only the proofs of the FTA use; the other theorems it proved were not requested. The architecture was designed around this property.

For the FTA, this property is the goal of the current work: GL is to find the lemmas of the FTA itself (\cref{sec:discovery}). Whether conjecture generation, filtering, proving and compression scale to this task remains to be shown.

\section{Availability}
\label{sec:availability}

\begin{sloppypar}
The code, the MPL definitions, the configuration files and the instructions to reproduce the runs of this paper are available at \href{https://github.com/Generative-Logic/GL}{\texttt{github.com/Generative-Logic/GL}}, release v0.13.0, commit \texttt{d666af87ce0bc660ef3b41853f95afd648fab083}. The proofs of the full run of \cref{sec:case_study}, as HTML pages and as a Lean~4 project, are in an archive attached to this release, with the SHA-256 hash of every file. The release is archived on Zenodo under \href{https://doi.org/10.5281/zenodo.23799906}{\texttt{doi:10.5281/zenodo.23799906}}. GL is available under the GNU Affero General Public License, version~3, or under a commercial license.
\end{sloppypar}

\section*{List of Abbreviations}

\begin{description}
    \item[AI] Artificial Intelligence
    \item[ASIC] Application-Specific Integrated Circuit
    \item[CAS] Computer Algebra System
    \item[CE] Counterexample
    \item[FPGA] Field-Programmable Gate Array
    \item[FTA] Fundamental Theorem of Arithmetic
    \item[GL] Generative Logic
    \item[GPU] Graphics Processing Unit
    \item[HTML] HyperText Markup Language
    \item[LB] Logic Block
    \item[LE] Logical Entity
    \item[LLM] Large Language Model
    \item[MMU] Memory Management Unit
    \item[MPL] Mathematical Programming Language
    \item[MPU] Mathematical Processing Unit
    \item[RAM] Random-Access Memory
    \item[SAT] Boolean Satisfiability
    \item[SMT] Satisfiability Modulo Theories
    \item[SRAM] Static Random-Access Memory
    \item[SSD] Solid-State Drive
\end{description}

\bibliographystyle{plainnat}
\bibliography{generative_logic}

\end{document}